\documentclass[structabstract]{aa}
\usepackage{txfonts}
\usepackage[dvips]{color}
\usepackage{graphicx}
\usepackage{natbib}
\bibpunct{(}{)}{;}{a}{}{,}
\usepackage{captcont}
\usepackage{subfig}
\usepackage{hyperref}
\begin{document}

\title{Outflow forces of low mass embedded objects in Ophiuchus: a quantitative comparison of analysis methods}
\titlerunning{Outflow forces: A quantitative comparison of analysis methods}
\author{N. van der Marel\inst{1} 
\and L.E. Kristensen\inst{1,2}
\and R. Visser\inst{3}
\and J.C. Mottram\inst{1}
\and U.A. Y{\i}ld{\i}z\inst{1}
\and E.F. van Dishoeck\inst{1,4}
}
\institute{Leiden Observatory, Leiden University, P.O. Box 9513, 2300 RA Leiden, the Netherlands
\and Harvard-Smithsonian Center for Astrophysics, 60 Garden Street, Cambridge, MA 02138, USA
\and Department of Astronomy, University of Michigan, 500 Church St., Ann Arbor, MI 48109-1042, USA
\and Max-Planck-Institut f\"{u}r Extraterrestrische Physik, Giessenbachstrasse 1, 85748 Garching, Germany
}
\date{\emph{(accepted in Astronomy \& Astrophysics, May 27 2013)}}

\abstract{The outflow force of molecular bipolar outflows is a key parameter in theories of young stellar feedback on their surroundings. The focus of many outflow studies is the correlation between the outflow force, bolometric luminosity and envelope mass. However, it is difficult to combine the results of different studies in large evolutionary plots over many orders of magnitude due to the range of data quality, analysis methods and corrections for observational effects such as opacity and inclination.}{We aim to determine the outflow force for a sample of low luminosity embedded sources. We will quantify the influence of the analysis method and the assumptions entering the calculation of the outflow force.}{We use the James Clerk Maxwell Telescope to map $^{12}$CO $J$=3--2 over 2$\arcmin\times2\arcmin$ regions around 16 Class I sources of a well-defined sample in Ophiuchus at 15$\arcsec$ resolution. The outflow force is then calculated using seven different methods differing e.g. in the use of intensity-weighted emission and correction factors for inclination. Two well studied outflows (HH~46 and NGC1333 IRAS~4A) are added to the sample and included in the comparison.}{The results from the analysis methods differ from each other by up to a factor of 6, whereas observational properties and choices in the analysis procedure affect the outflow force by up to a factor of 4. Subtraction of cloud emission and integrating over the remaining profile increases the outflow force at most by a factor of 4 compared to line wing integration. For the sample of Class I objects, bipolar outflows are detected around 13 sources including 5 new detections, where the three non-detections are confused by nearby outflows from other sources. New outflow structures without a clear powering source are discovered at the corners of some of the maps.}{When combining outflow forces from different studies, a scatter by up to a factor of 5 can be expected. Although the true outflow force remains unknown, the separation method (separate calculation of dynamical time and momentum) is least affected by the uncertain observational parameters. The correlations between outflow force, bolometric luminosity and envelope mass are further confirmed down to low luminosity sources.
}

\keywords{ISM: jets and outflows -- ISM: molecules -- stars: protostars -- stars: low-mass -- stars: circumstellar matter -- submillimeter: ISM}

\maketitle

\section{Introduction}
Molecular outflow activity begins early during the formation process of low-mass protostars and is therefore a tracer of active star formation \citep{bachiller1999,arce2006}. Outflows actively contribute to the core collapse and formation of the star by carrying away angular momentum that would otherwise prevent accretion \citep{bachiller1996,hogerheijde1998}. The strength of an outflow is measured by the outflow force $F_{\rm CO}$: the ratio of the momentum and the dynamical age of the outflow \citep{bachiller1999}. It is a measure of the rate at which momentum is injected into the envelope by the protostellar outflow \citep{downes2007}. The outflow force has been used to understand the driving mechanism of outflows. The correlation between the bolometric luminosity $L_{\rm bol}$ and outflow force suggests that a single mechanism is responsible for driving the outflow, at least partly related to accretion \citep{cabrit1992}. A second correlation between $F_{\rm CO}$ and $M_{\rm env}$ \citep{bontemps1996} underlines the evolutionary properties of this parameter, suggesting that the outflow activity declines during the later stages of the accretion phase though this is more difficult to separate from initial conditions \citep{hatchell2007}. Due to these correlations, the outflow force is a very important parameter for understanding the physical processes underlying the early phases of star formation. 

Outflow forces have been derived for many embedded sources over the past 30 years using various methods and with continuous increase of observational resolution and $S/N$, resulting in a very inhomogeneous data set. Outflow properties for low-mass protostars have been compared in population studies of entire star forming regions and in different phases of evolution, ranging from the deeply embedded Class 0 to the T Tauri Class II stage \citep{cabrit1992,bontemps1996,richer2000,arce2006,hatchell2007}. A central focus of many of these studies are the evolutionary trend plots of $F_{\rm CO}$ versus $L_{\rm bol}$ and $F_{\rm CO}$ versus $M_{\rm env}$ as described above. However, the results of these studies cannot usually be combined into a single dataset including all nearby clouds due to the range of data quality, the different methods used to determine outflow properties and the different corrections for observational effects, such as opacity and inclination, that have been applied. For individual targets, different methods have been shown to result in different values of the outflow force up to an order of magnitude \citep{hatchell2007}. Comparisons in the value of the outflow force with different CO transitions (e.g. $J$=1--0, 2--1, 3--2, 6--5) show similar results \citep{kempen2009hh46,nakamura2011,yildiz2012} . Although some comparison studies with outflow models exist (Cabrit \& Bertout 1992, CB92 hereafter; Downes \& Cabrit 2007, DC07 hereafter), the influence of the analysis method itself and the assumptions entering the calculation have never been quantified systematically against a single set of observational data. 

The $\rho$ Oph molecular cloud complex \citep[$d\sim 120\pm4$ pc;][]{loinard2008} is one of the nearest and best studied star forming regions \citep{lada1984,greene1994}. Tens of young stellar objects (YSOs) have been identified, especially in the main filament L1688 (Wilking et al. 2008, and references therein).  Due to the clustering of star formation in Ophiuchus, the YSOs are very close together and their outflows sometimes overlap along the line of sight \citep[e.g.][]{bussmann2007}. Class I outflows in Ophiuchus have been studied extensively and their properties have been obtained for many sources \citep{bontemps1996,sekimoto1997,kamazaki2001,kamazaki2003,ceccarelli2002,boogert2002,bussmann2007,gurney2008,zhang2009,nakamura2011}. Additional studies of YSOs in Ophiuchus from earlier data, including sources with outflows, are summarized in \citet{wilking2008}. It has been suggested that star formation in Ophiuchus was triggered by supernovae, ionization fronts and winds from the Upper Scorpius OB association, located to the west of the Ophiuchus cloud \citep{preibisch1999,nutter2006} and started in the denser northwestern L1689 region \citep{zhang2009}.

In a recent study by \citet{kempen2009}, a number of YSOs in Ophiuchus were classified as so-called `Stage 1' sources, using a new classification based on $M_{\rm disk}$, $M_{\rm env}$ and $M_{\rm star}$. Since this classification relies on physical rather than observational parameters \citep[c.f. Lada classification, with spectral slope $\alpha_{\rm IR}$,][]{lada1984}, this sample represents a well-defined embedded YSO sample in this cloud and outflow activity is expected to be detected from every source. For about half of these sources bipolar outflows were identified previously, but not with the high spatial resolution of 15$\arcsec$ that is available with the HARP array receiver \citep{buckle2009} at the James Clerk Maxwell Telescope (JCMT). High-resolution $^{12}$CO $J$=3--2 spectral maps offer the means to study the occurence, origin and properties of the outflows in more detail than previous studies. Therefore, this data set is perfect to systematically study and quantify the effect of different analysis methods. The sources are also part of the \emph{Herschel} key programs ``Water in Star-forming Regions with Herschel" (WISH) and ``Dust, Ice and Gas In Time" (DIGIT) \citep[Green et al.\ (in prep.)]{vandishoeck2011}: complementary data on feedback based on far infrared line emission will become available in the following years for comparison.

This paper presents the results of a comparison between analysis methods of the outflow force with CO spectral maps, applied to 13 sources in Ophiuchus and two additional, well-studied outflows: HH~46 \citep{kempen2009hh46} and NGC1333 IRAS~4A \citep{yildiz2012}. The outline of this paper is as follows. In Sect. 2, the details of the observations and basic data reduction are discussed. Section 3 presents the spectral profiles and contour maps of the integrated line wings, indicating the outflow lobes. In Sect. 4, the results of the various outflow force analysis methods are presented and discussed. Section 5 discusses the comparison of outflow force values with those found in the literature, the trends of the outflow parameters and its implications for evolutionary models and the comparison of outflow morphologies with observed disks. The conclusions of the paper are given in Sect. 6.

\section{Observations}
\subsection{Sample selection}
\begin{figure*}[tp]
\begin{center}
\includegraphics[width=15cm,trim=0 -50 0 0]{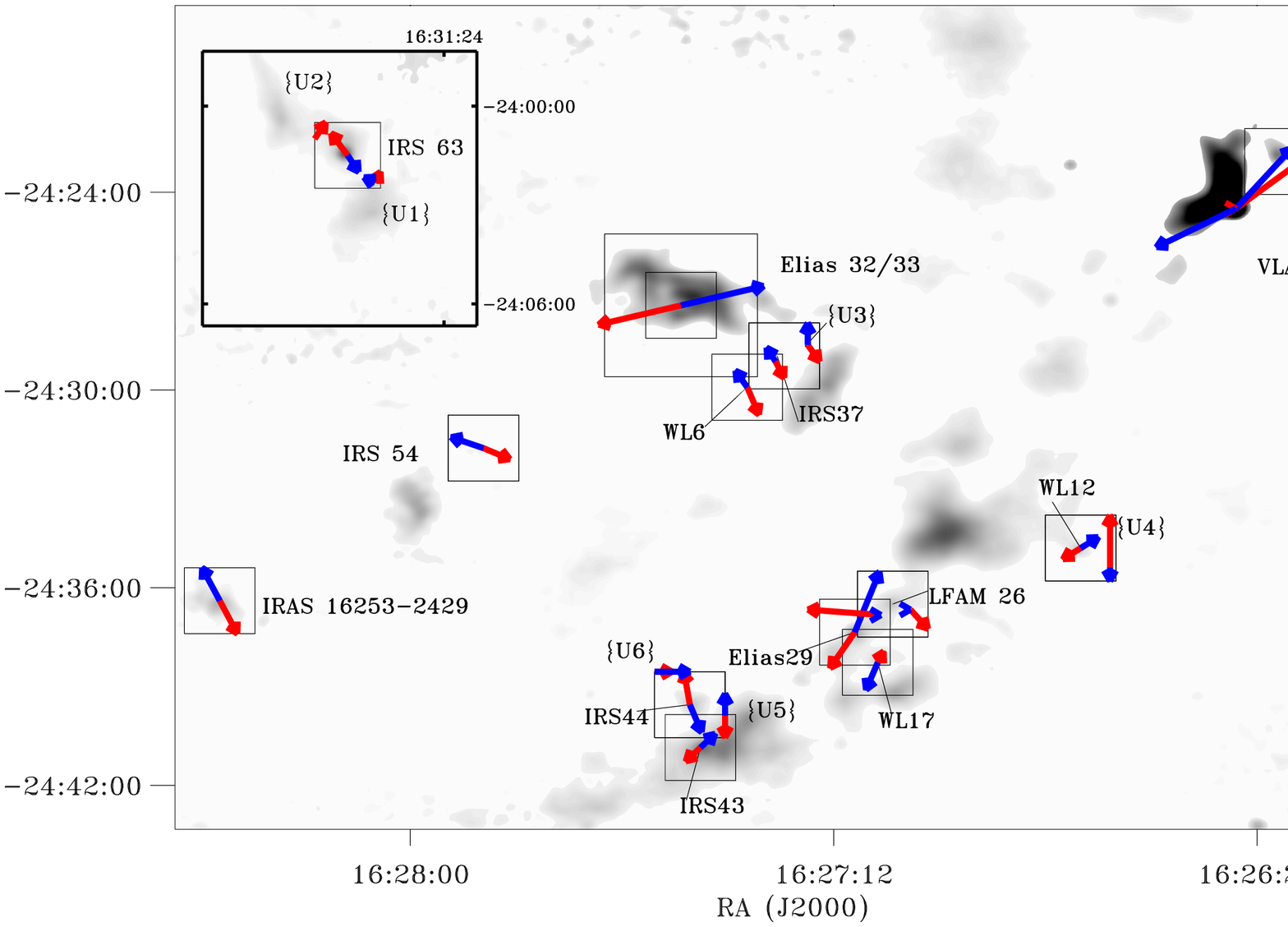}
\caption{The L1688 core in Ophiuchus and the region around IRS~63 (inset upper left corner). The background shows the 850 $\mu$m SCUBA map \citep{johnstone2000,difrancesco2008}. The locations of all observed HARP maps are indicated by squares, labeled by the sources within. The blue and red arrows indicate the direction and observed extent of the blue and red outflow of that source, except for the extent of VLA 1623, which was taken from \citet{yu1997} who observed the region centered at VLA~1623. In crowded regions, lines connect the source names to the origin location of the outflow. Newly detected outflow structures without a driving source are marked with Us. }
\label{bigmap}
\end{center}
\end{figure*}

The 16 `Stage 1' sources identified by \citet{kempen2009} were selected for the sample of this outflow study, out of a sample of 41 sources with potential Class I classification ($\alpha_{\rm 2-24\mu{}m} >$ 0.3 and $T_{\rm bol} <$ 650 K). Figure \ref{bigmap} presents an overview of our sample, the basic properties of which are shown in Table \ref{sample}. IRS~46 is classified as Stage 2, but an outflow was also detected for this source in this study. The $L_{\rm bol}$ for this sample ranges from 0.04 to 14.1 $L_{\odot}$. Except for IRS~63, all sources are located in the L 1688 ridge. 

Figure \ref{bigmap} shows an overview of the bipolar outflows and their relative positions in the L1688 core in Ophiuchus, where each square represents a spectral map and the red and blue arrows show the direction and extent of the red and blue outflows. VLA~1623 is added as well, based on figures in \citet{yu1997}. The background shows the 850 $\mu$m SCUBA map as published by \citet{johnstone2000} and \citet{difrancesco2008}. IRS~63 lies outside the borders of this map and is presented as inset. 

\begin{center}
\begin{table*}[ht]
\caption{Sample of Stage 1 sources in Ophiuchus.}
\label{sample}
\centering
\begin{tabular}{llllrrr}
\hline
\hline
Source &\multicolumn{2}{c}{Coordinates (J2000)}& Alternative names&$L_{\rm bol}$\tablefootmark{a}&$M_{\rm env}$\tablefootmark{a}&$i$\tablefootmark{c}\\
&RA&Dec&&($L_{\odot}$)&(10$^{-2}$ $M_{\odot}$)&($\degr$)\\
\hline
GSS~30-IRS1&16:26:21.4&$-$24:23:04.1&GSS~30, Elias~21&3.3\phantom{5}&20.5\phantom{\tablefootmark{b}}&-\\
GSS~30-IRS3&16:26:21.7&$-$24:22:51.4&LFAM~1&0.83&17.1\phantom{\tablefootmark{b}}&-\\
WL~12&16:26:44.0&$-$24:34:48.0&GY~111&3.4\phantom{5}&4.6\phantom{\tablefootmark{b}}&30\\
LFAM~26&16:27:05.3&$-$24:36:29.8&GY~197, CRBR~2403.7&0.04&4.5\phantom{\tablefootmark{b}}&70\\
WL~17&16:27:07.0&$-$24:38:16.0&GY~205&0.67&4.0\phantom{\tablefootmark{b}}&50\\
Elias~29&16:27:09.6&$-$24:37:21.0&WL15, GY~214&14.1\tablefootmark{b}&4.0\tablefootmark{b}&50\\
IRS~37&16:27:17.6&$-$24:28:58.0&GY~244&0.38&1.2\phantom{\tablefootmark{b}}&50\\
WL~3&16:27:19.3&$-$24:28:45.0&GY~249&0.46&2.9\phantom{\tablefootmark{b}}&-\\
WL~6&16:27:21.8&$-$24:29:55.0&GY~254&0.85&$<$0.4\phantom{\tablefootmark{b}}&50\\
IRS~43&16:27:27.1&$-$24:40:51.0&GY~265, YLW~15&1.0\phantom{5}&17.1\phantom{\tablefootmark{b}}&10\\
IRS~44&16:27:28.3&$-$24:39:33.0&GY~269&1.1\phantom{5}&8.0\phantom{\tablefootmark{b}}&30\\
IRS~46&16:27:29.7&$-$24:39:16.0&GY~274&0.19&3.3\phantom{\tablefootmark{b}}&50\\
Elias~32&16:27:28.6&$-$24:27:19.8&IRS~45, VSSG~18&0.5\phantom{5}&41.9\phantom{\tablefootmark{b}}&-\\
Elias~33&16:27:30.1&$-$24:27:43.0&IRS~47, VSSG~17&1.2\phantom{5}&28.2\phantom{\tablefootmark{b}}&70\\
IRS~54&16:27:51.7&$-$24:31:46.0&GY~378&0.78&3.1\phantom{\tablefootmark{b}}&50\\
IRAS~16253--2429&16:28:21.6&$-$24:36:23.7&&0.06&10.3\phantom{\tablefootmark{b}}&70\\
IRS~63&16:31:35.7&$-$24:01:29.5&GWAYL~4&1.0\tablefootmark{b}&30.0\tablefootmark{b}&30\\
\hline
\end{tabular}
\tablefoot{
\tablefoottext{a} Parameters taken from \citet{kempen2009}, unless indicated otherwise. 
\tablefoottext{b} From \citet{kristensen2012}. 
\tablefoottext{c} Estimates from this work, see Sect. \ref{recipe}.
}
\end{table*}
\end{center}

\subsection{$^{12}$CO maps of Ophiuchus}
\label{sampleselection}
All sources were observed in the $J$=3--2 line of $^{12}$CO (345.796 GHz) with the HARP array receiver at the James Clerk Maxwell Telescope (JCMT) as part of program M08AN05. A bandwidth of 250 MHz was used with the ACSIS backend, resulting in a native spectral resolution of 0.026 km s$^{-1}$. The observations were carried out in June 2008 under weather conditions with an atmospheric opacity $\tau_{\rm 225GHz}$ ranging from 0.08 to 0.12. For analysis of the line wings, the spectra are typically binned to 0.1 km s$^{-1}$, giving $\sigma_{\rm rms}$ = 0.2 K. 
The HARP array consists of 16 receivers arranged in a 4$\times$4 pattern, mapping the sources over 2$\arcmin\times2\arcmin$ to capture the full extent of the outflows with 15$\arcsec$ spatial resolution. The maps were made in the jiggle observing mode and resampled with a pixel size of 7$\farcs$5. A position switch of 60$\arcmin$ or 150$\arcmin$ was used. For one night, the reference off position accidentally coincided with emission, resulting in negative absorption troughs in the spectra in the maps of Elias~32 and Elias~33, IRS~37, IRS~43, IRS~44, LFAM~26, WL~17 and IRS~54. This has, however, no influence on the study of line wings. 
The Class 0 source IRAS 16293--2422 was used as a line calibrator. The main-beam efficiency was 0.63\footnote{\url{http://www.jach.hawaii.edu/JCMT/spectral_line/}\\ \indent \url{General/status.html}} and pointing errors were within 2$\arcsec$. Two receivers were broken at the time of observation (H03 and H014), resulting in lack of data in the south-east and the north-north-west corner of each map. Some sources are separated by less than 30$\arcsec$ and were therefore mapped in a single image (Elias~32 + Elias~33; IRS~37 + WL~3; GSS30-IRS1 + GSS30-IRS3; and IRS~44 + IRS~46). Only for the last case the two sources could be analyzed separately.
In addition, Elias~33 was observed in September 2007 with HARP in the raster observing mode. This source was mapped in a 4$\arcmin\times4\arcmin$ region and resampled to a pixel size of 12$\arcsec$. Atmospheric opacity was 0.05. The rms noise $\sigma_{\rm rms}$ for the rebinned spectra is again 0.2 K in 0.1 km s$^{-1}$ bins.

\subsection{$^{13}$CO spectra}
In order to determine the optical depth of the line wings of the outflows additional $^{13}$CO $J$=3--2 (330.58797 GHz) spectra were taken for two of the strongest outflows (Elias~29 and IRS~44) at the central positions. The HARP array receiver at the JCMT was used in the same setup as the $^{12}$CO observations as single pointings instead of jiggle maps. The observations were carried out in March 2011 as part of program M10BN05 under weather conditions with an opacity $\tau_{\rm 225GHz}$ of $\sim$0.11. The rms noise $\sigma_{\rm rms}$ for the reduced spectra is  0.05 K in 0.4 km s$^{-1}$ bins.
Although all 16 receivers recorded a spectrum, only the spectrum at the central position showed line wings and so could be used for the optical depth derivation of the $^{12}$CO line wings. This will be discussed further in Sect. \ref{sect-opticaldepth}.

\subsection{$^{12}$CO maps of HH46 and IRAS4A}
For the outflow analysis comparison, $^{12}$CO $J$=3--2 datasets published in \citet{kempen2009hh46} and \citet{yildiz2012} of HH~46 and IRAS~4A were used, respectively. The HH~46 observations were taken at the Atacama Pathfinder EXperiment (APEX), regridded to a map of $2\arcmin\times2\arcmin$ and have a typical rms noise of 0.6 K in 0.2 km s$^{-1}$ bins. The observations of IRAS~4A were taken in jiggle mode with the HARP array receiver at the JCMT, regridded to a map of $4\arcmin\times4\arcmin$ have a typical rms noise of 0.15 K in 0.1 km s$^{-1}$ bins. For more details on these observations, see the original references. 

\section{Results}
\subsection{Spectral profiles}
\begin{figure}[tp]
\includegraphics[scale=0.5]{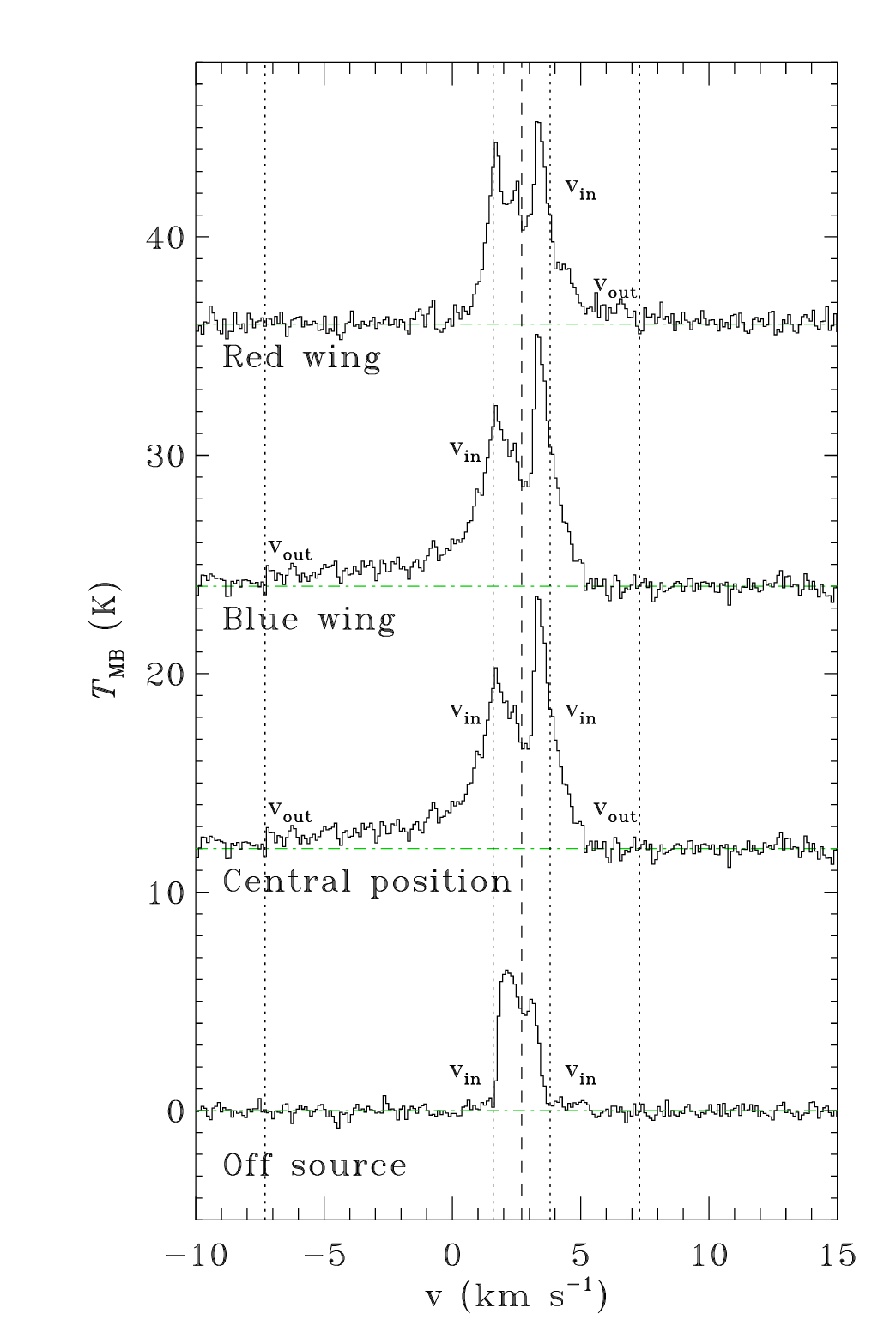}
\caption{$^{12}$CO 3--2 spectra toward IRS~63, from top to bottom: one red outflow lobe position (15$\arcsec$,15$\arcsec$), one blue outflow lobe position (0$\arcsec$, $-$7$\farcs$5), the central position (0$\arcsec$,0$\arcsec$) and one off-source position ($-$42$\farcs$5,20$\arcsec$). The integration limits are marked by dotted lines, baselines by dash-dotted lines and the source velocity by a dashed line.}
\label{example}
\end{figure}
$^{12}$CO $J$=3--2 is detected towards all sources, showing broad line profiles centered around $\varv_{\rm LSR}\sim$4 km s$^{-1}$. An example set of spectra is shown in Fig. \ref{example} for IRS~63. The spectral profiles consist of red/blue-shifted wing emission from the outflow, central emission peaks from cloud and envelope material or foreground layers and narrow absorption features caused by self-absorption.  
At the outflow positions, broad line wings are detected, up to +15 km s$^{-1}$ in the red and $-$10 km s$^{-1}$ in the blue. 

\subsection{Outflow maps}
\begin{figure*}[tp]
\includegraphics[width=16cm]{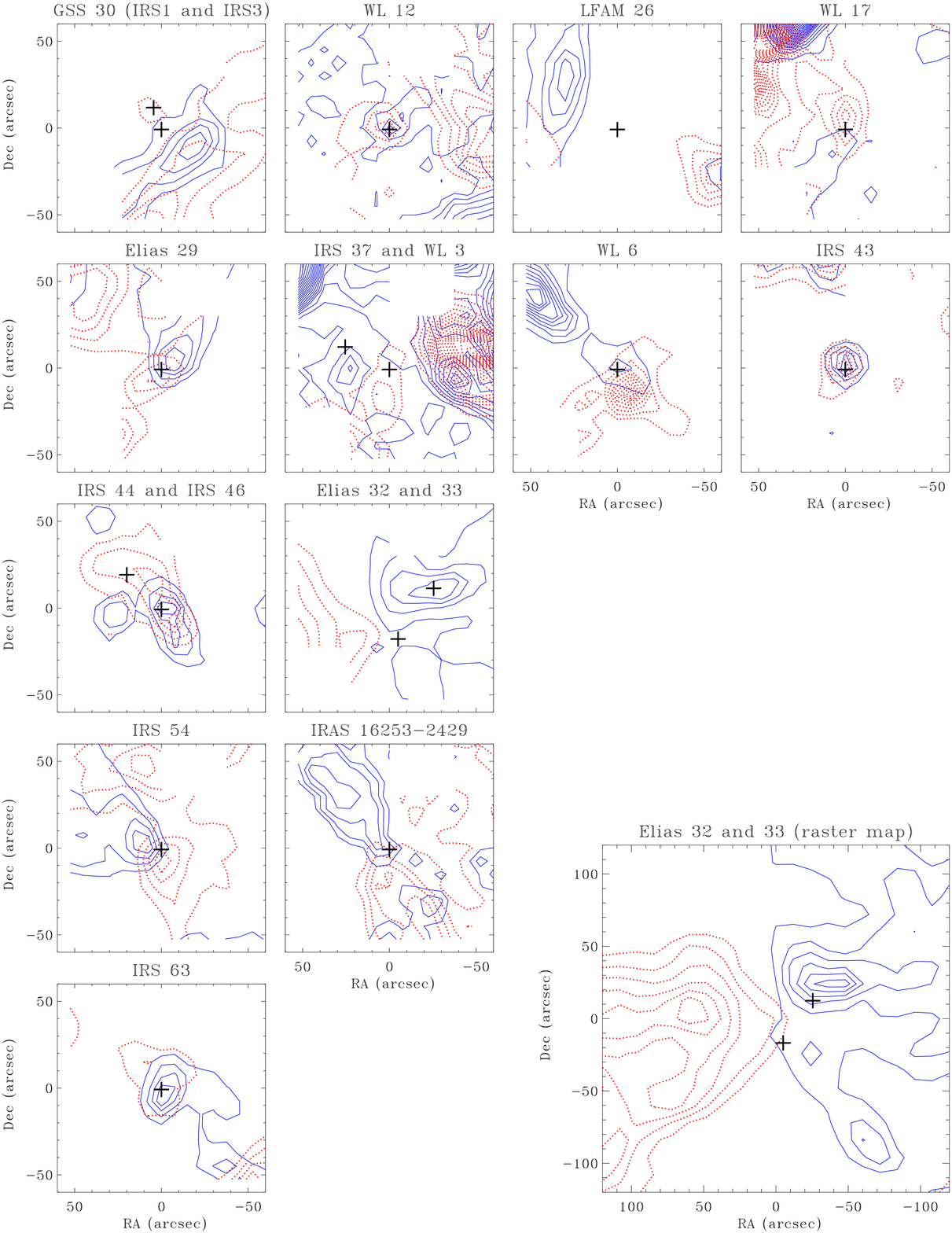}
\caption{Outflow maps of all sources. Contours are drawn based on the integrated intensities of the line wings. Contours are drawn at 20, 40, 60, 80 and 100\% of the peak value of the integrated intensity of the target source. The source positions are marked by pluses. Two receivers were broken at the time of observation, resulting in lack of data in the south-east  and north-north-west corner of each map. In maps with Elias 32 and 33, Elias 32 is north west of Elias 33. In the GSS 30 map, GSS~30-IRS3 is north of GSS~30-IRS1. In the IRS~37 map, WL~3 is east of IRS~37. In the IRS~44 map, IRS~46 is north east of IRS~44.}
\label{outflowmaps}
\end{figure*}

Outflow maps were made of the integrated CO wing emission in blue and red velocity intervals. The integration limits were chosen as follows. For each position, the outer velocity, $\varv_{\rm out}$, is defined as the velocity where the emission is still above the 1$\sigma$ rms level in 0.1 km s$^{-1}$ bins.  
The inner velocity limits, $\varv_{\rm in}$, in each map are defined as the $\varv_{\rm out}$ in an off-source $^{12}$CO spectrum (see Fig. \ref{example} for the example of IRS~63). The blue wing integration limits typically range from $-$6 km s$^{-1}$ to +1 km s$^{-1}$, and for the red wing from 6 km s$^{-1}$ to 12 km s$^{-1}$ (see Fig. \ref{overviewspect}). Contour maps of the integrated outflow emission are presented in Fig. \ref{outflowmaps}.

The maps show outflow activity in most of the sources. Infrared source positions are at (0$\arcsec$,0$\arcsec$) or marked when there are multiple sources. Most outflows fit in the 2$\arcmin\times2\arcmin$ region, except for Elias~32/33, Elias~29, LFAM~26 and IRAS~16253--2429, although the outflow direction can still be recognized. Elias~32/33 is also observed in the raster mode in 4$\arcmin\times4\arcmin$ maps, covering a larger part of the outflow (see lower right panel in Fig. \ref{outflowmaps}). 

Bipolar outflows are seen for 13 sources (see Table \ref{outflowstatus}), although nearby outflows complicate the view in several cases. In the Elias~29 map, emission from the east lobe of LFAM~26 is seen in the upper right corner, while part of the blue lobe of Elias~29 falls in the LFAM~26 map and the red lobe in the WL~17 map. A full coverage of these two outflows can be found in \citet{bussmann2007} and \citet{nakamura2011}. The red lobe of Elias~32/33 extends all the way to IRS~54 and the blue lobe is wide enough to show close to WL~6 \citep{nakamura2011}. The lobes of IRS~44 extend into the IRS~43 map. In the map containing GSS~30, the outflow emission is completely dominated by the 15$\arcmin$ long outflow extending from VLA~1623, a Class 0 source at 16:26:26.26, $-$24:24:30.01 \citep{andre1993,dent1995,yu1997}. A high-velocity bullet \citep{bachiller1999} at 28 km s$^{-1}$, not identified in earlier studies of VLA~1623, was discovered 50$\arcsec$ south of GSS~30 (see Fig. \ref{GSS30highvelspec}). \citet{jorgensen2009} resolved an outflow lobe for GSS~30-IRS1 in interferometry HCO$^+$ $J$=3--2 observations, tracing the inner outflow cone, but our CO observations do not reveal this outflow. In the case of Elias~29 and Elias~32/33, the red and blue lobes overlap with the pixels from the broken receivers, thus possibly underestimating the total outflow mass. 

In the maps of Elias~32/33 and IRS~37/WL~3, only one set of outflow lobes was resolved: for this study they were assigned to Elias~33 and IRS~37, respectively. In the map of IRS~44, the Class II source IRS~46 is located 20$\arcsec$ north-east of IRS~44. Based on the sudden change of shape of the line wings, the spectral map was cut in two, assigning one half to IRS~46 and one half to IRS~44, each with their own bipolar outflow. In the maps containing IRS~37, IRS~44, IRS~63 and WL~12 new outflow structures were discovered which could not be assigned to an IR source. These structures were assigned U1, U2, etc. and are discussed further in Appendix \ref{Us}.

Based on the outflow maps in Fig. \ref{outflowmaps}, an outflow status can be assigned to each source, which is listed in Table \ref{outflowstatus}. Four new outflows were identified and nine were confirmed from previous observations. 

\begin{center}
\begin{table}[tp]
\caption{Outflow status for the 16 sources in this sample.}
\label{outflowstatus}
\centering
\begin{tabular}{lcc}
\hline
\hline
Source & Outflow status\tablefootmark{a} &References\\
\hline
GSS~30$-$IRS1\tablefootmark{b}&C&\ldots\\
GSS~30$-$IRS3\tablefootmark{b}&C&\ldots\\
WL~12&B&1\tablefootmark{c,d}\\
LFAM~26&B&2,3\\
WL~17&B&New\\
Elias~29&B&1,2,3,4\\
IRS~37&B&New\\
WL~3&C&\ldots\\
WL~6&B&1,4\\
IRS~43&B&1\\
IRS~44&B&1,4\\
IRS~46&B&New\\
Elias~32/33\tablefootmark{e}&B&3,5\\
IRS~54&B&New\\
IRAS~16253--2429&B&New\\
IRS~63&B&1\tablefootmark{c,d}\\
\hline
\end{tabular}
\tablefoot{
\tablefoottext{a} Outflow status: $B$ for bipolar outflow and $C$ for confusion because of the presence of nearby sources. 
\tablefoottext{b} Outflow confusion with VLA 1623.
\tablefoottext{c} IRS~63 is named L1709B (name of the core) or 16285-2355 in \citet{bontemps1996}.
\tablefoottext{d} IRS~63 and WL~12 were not recognized as bipolar in \citet{bontemps1996}.
\tablefoottext{e} Due to the small separation of Elias~32 and 33, it is not clear to which source the outflow belongs. Elias~33 is assigned as the driving source in this study.
}
\tablebib{
(1) \citet{bontemps1996}, (2) \citet{bussmann2007}, (3) \citet{nakamura2011}, (4) \citet{sekimoto1997}, (5) \citet{kamazaki2003}.
}
\end{table}
\end{center}

\subsection{Optical depth}
\label{sect-opticaldepth}
\begin{figure}[tp]
\includegraphics[scale=0.4]{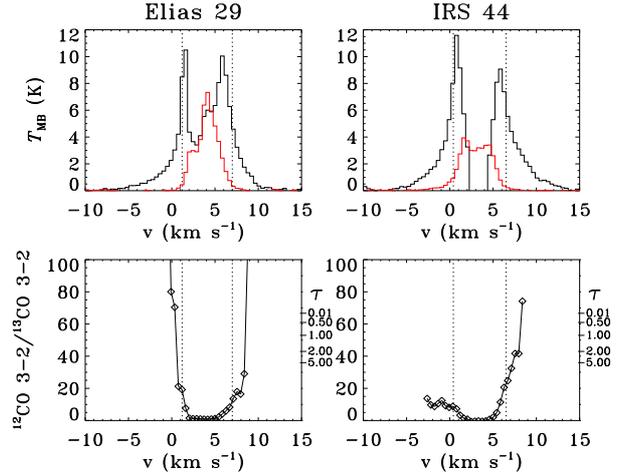}
\caption{$^{12}$CO 3--2 and $^{13}$CO 3--2 spectra of the central position of Elias~29 (left) and IRS~44 (right). The top figure shows the spectra, binned to 0.42 km s$^{-1}$, the bottom figure shows the $T_{\rm MB}$ ratio of $^{12}$CO 3--2/$^{13}$CO 3--2 and the corresponding optical depth $\tau$ on the right axis. The dotted lines show the integration limits, indicating the velocities $\varv_{\rm in}$ where the line wings start.}
\label{opticaldepth}
\end{figure}

The optical depth, $\tau$, is obtained from the line ratio of $^{12}$CO $J$=3--2 and its isotopolog $^{13}$CO $J$=3--2 for the central position of the outflows with the strongest line wings, Elias~29 and IRS~44. In Fig. \ref{opticaldepth}, spectra of $^{12}$CO $J$=3--2 and $^{13}$CO $J$=3--2 at the central positions are shown, binned to 0.42 km s$^{-1}$, together with the ratio of $^{12}$CO $J$=3--2/$^{13}$CO $J$=3--2. The optical depth for the line wings is derived assuming that the two species have the same excitation temperature and that the $^{13}$CO line wing is optically thin, using \citep{goldsmith1984}:
\begin{equation}
\frac{I(^{12}{\rm CO})}{I(^{13}{\rm CO})} = \frac{1-e^{-\tau_{12}}}{1-e^{-\tau_{13}}} \approx \frac{1-e^{-\tau_{12}}}{\tau_{12}}\times R
\end{equation}

with $R = \tau_{12}/\tau_{13}$ the abundance ratio $^{12}$CO/$^{13}$CO \citep[a ratio of 65 is assumed here, following][]{vladilo1993}. The resulting optical depths of $^{12}$CO as a function of velocity are shown on the right-hand axes of the lower panels of Fig. \ref{opticaldepth}. High optical depths $>$2 are found at velocities very close to the source velocity $\varv_{\rm source}$ implying that the central velocities are optically thick and become optically thinner away from the line center where the outflow emission starts. The only exception is the blue wing of IRS~44, which remains optically thick for 3 km s$^{-1}$ beyond $\varv_{\rm in}$. Quantitatively, the optical depth implies that the mass of the blue lobe at this single position is underestimated by a factor of 6 if the wing emission is assumed to be optically thin. At the other positions with $^{13}$CO measurements, the blue wing is still optically thin, so the total blue mass is not expected to increase more than a factor of 2 due to optical depth effects. Since the uncertainties on the outflow force are within a factor of a few, the impact of optical depth is negligible. Therefore outflow emission is assumed to be optically thin for all sources. 

\section{Outflow analysis}
\label{sectioncompare}
\subsection{Outflow parameters}
The main physical parameters of the outflow are the mass, $M$, the velocity, $\varv_{\rm CO}$ (defined as $|\varv_{\rm out}-\varv_{\rm source}|$, with $\varv_{\rm out}$ the outer velocity of the outflow emission at a given location), and the projected size of the lobe $R_{\rm lobe}$, for both the blue and red lobe. The mass is calculated from the integrated intensity of the $^{12}$CO line assuming an excitation temperature of 50 K \citep[e.g.][]{yildiz2012} and the procedure explained in Appendix \ref{derivemass}. With these basic physical quantities the outflow force, $F_{\rm CO}$, can be derived as 
\begin{equation}
F_{\rm CO} = \frac{M\varv_{\rm CO}^2}{R_{\rm lobe}}
\end{equation}

The velocity, $\varv_{\rm CO}$, often called $\varv_{\rm max}$, is not a well-constrained parameter: due to the inclination and the shape of the outflow, often following a bow shock including forward and transverse motion, the measured velocities along the line of sight are not necessarily representative of the real velocities driving the outflow \citep{cabrit1992}. This is the main reason why various methods have been developed to derive the outflow force, where a correction factor is often applied to compensate for these effects. 

\subsection{Analysis methods}
\label{recipe}
Seven different methods for deriving the outflow force are analytically described in this section, with references to the paper where the method was originally presented. The constant $K$ refers to Equation \ref{convmass}. The following general recipe applies to each method for the blue and red lobe separately. The outflow parameters are calculated by adding up the absolute values for the blue and red lobe.
\begin{enumerate}
\item For each pixel, $j$, the outer velocity $\varv_{{\rm out},j}$ is defined as the outer velocity where the emission is still above the 1$\sigma$ rms level.
\item The inner velocity, $\varv_{\rm in}$, in each map is defined as the velocity in the average off-source $^{12}$CO spectrum where the emission is still above 5\% of the peak emission $T_{\rm peak}$. $\varv_{\rm in}$ and $\varv_{{\rm out},j}$ define the integration limits for each line wing, the same $\varv_{\rm in}$ is used for each pixel.
\item The source velocity, $\varv_{\rm source}$, is determined using a high-density tracer such as HCO$^+$ or a minor isotopologue with a non-self-absorbed profile. $\varv' = \varv-\varv_{\rm source}$ is defined as the relative velocity.
\item Only pixels that are contributing to the outflow of the target are included to avoid confusion with nearby outflows. 
\item The inclination angle, $i$, is estimated from the morphology of the contour map to be pole on (10$\degr$), inclined  (30$\degr$ or 50$\degr$) or in  the plane of the sky  (70$\degr$), using Fig. 1-4 in \citet{cabrit1990}. 
\item $R_{\rm lobe}$ is measured as the projected length of the outflow lobe.
\item The conversion from integrated intensity to mass is a multiplication with constant $K$ as defined in Appendix \ref{derivemass}.
\end{enumerate}

\begin{center}
\begin{table}[tp]
\caption{Correction factors.}
\label{corrections}
\begin{center}
\begin{tabular}{lrrrrl}
\hline
\hline
$i$($\degr$)&10&30&50&70&Ref.\\
\hline
$c_1$&0.28&0.45&0.45&1.1\phantom{5}&1,2\\
$c_2$&1.6\phantom{5}&3.6\phantom{5}&6.3\phantom{5}&14\phantom{.55}&1\\
$c_3$\tablefootmark{a}&1.2\phantom{5}&2.8\phantom{5}&4.4\phantom{5}&7.1\phantom{5}&3\\
\hline
\end{tabular}
\end{center}
\tablebib{
(1) \citet{cabrit1990}, (2) CB92, (3) \citet{downes2007}. }
\tablefoot{
\tablefoottext{a}The values are interpolated from the ratios in the third column of Table 6 of \citet{downes2007}, where $\alpha = 90-i$.
}
\end{table}
\end{center}

\subsubsection{$\varv_{\rm max}$ method (M1)}
M1 is one of the most common methods in the literature \citep[e.g.][]{cabrit1992,hogerheijde1998,downes2007,kempen2009outflow}. It assumes a constant $\varv'_{\rm max} = {\rm max}(\varv'_{{\rm out},j})$ throughout the spectral map. The derived outflow force, $F_{\rm obs}$, is multiplied by the inclination correction factors $c_1(i)$ (see Table \ref{corrections}) derived by \citet{cabrit1990,cabrit1992} for $\varv_{\rm CO}$ to obtain $F_{\rm CO}$:
\begin{equation}
F_{\rm CO} = c_1 \frac{K\left(\displaystyle\sum\limits_j \left[\int T(\varv){\mathrm d}\varv\right]_j \right) {\varv'}_{\rm max}^2}{R_{\rm lobe}} 
\end{equation}

\subsubsection{$\varv_{\rm spread}$ method (M2)}
In M2, $\varv_{{\rm out},j}$ is derived as a function of position \citep{yildiz2012}. The local energy $M\varv_{{\rm CO},j}^2$ is calculated for each position in the map with the local velocity. No correction factors are applied:
\begin{equation}
F_{\rm CO} = \frac{K\left(\displaystyle\sum\limits_j {\varv'}_{{\rm out},j}^2\times\left[\int T(\varv){\mathrm d}\varv\right]_j\right)}{R_{\rm lobe}}
\end{equation}

\subsubsection{$\langle\varv\rangle$ method (M3)}
M3 uses intensity-weighted velocities by including $\varv'$ in the integral \citep{cabrit1990,downes2007}. The derived outflow force is multiplied by the inclination correction factors $c_2(i)$ (Table \ref{corrections}) derived in \citet{cabrit1990} for $\langle\varv\rangle$ to obtain $F_{\rm CO}$:
\begin{equation}
F_{\rm CO} = c_2 \times K\left(\frac{\left\{\displaystyle\sum\limits_j \left[\int T(\varv')\varv'{\mathrm d}\varv'\right]_j\right\}^2}{R_{\rm lobe}\displaystyle\sum\limits_j \left[\int T(\varv'){\mathrm d}\varv'\right]_j}\right)  \\
\end{equation}

\subsubsection{Local method (M4)}
Using the kinematic structure of the outflow lobe, the local outflow force is calculated for each position by dividing the local energy $M\varv_{{\rm CO},j}^2$ by the projected distance $r_j$ between the position and source position \citep{lada1996,downes2007}. \citet{hatchell2007} use a combination of M4 and M2, using projected distances and local velocities. No correction factors are applied:
\begin{equation}
F_{\rm CO} = K\left(\displaystyle\sum\limits_j \frac{ \left[\int T(\varv'){\varv'}^2{\mathrm d}\varv'\right]_j}{r_j}\right) 
\end{equation}

\subsubsection{Perpendicular method (M5)}
Because a large amount of the outflow material is moving slowly and predominantly in transverse motion, DC07 developed a method to determine the dynamical age, $t_{\rm d}$, of the outflow, using the half-width of the outflow lobe, $W_{\rm lobe}$, rather than the projected length. The factor $\frac{1}{3}$ (see below) originates from the asymptotic expansion of the bowshock wings, $R\propto t^{1/3}$ (DC07 and references therein). This method resulted in much more accurate estimates of the outflow force for the modeled outflows, but can only be applied in observations of spatially well-resolved outflows. In addition, since the models describe Class 0 sources, the results may not be applicable for outflows of Class I sources which are known to have large opening angles \citep[e.g.][]{arce2006}:
\begin{equation}
F_{\rm CO} = K\left(\frac{\left\{\displaystyle\sum\limits_j \left[\int T(\varv')\varv'{\mathrm d}\varv'\right]_j\right\}^2}{\frac{1}{3}W_{\rm lobe} \displaystyle\sum\limits_j \left[\int T(\varv'){\mathrm d}\varv'\right]_j}\right) 
\end{equation}

\subsubsection{Annulus method (M6)}
In M6, the outflow force is calculated with a ``slice'' through the outflow \citep{bontemps1996}. The momentum within an annulus or ring centered on the source position with thickness $\Delta{r}$ (equal to the beam size) and radius $r$ is measured, and the dynamical age is determined with $\Delta{r}$ rather than $R_{\rm lobe}$. The derived outflow force $F_{\rm obs}$ is multiplied by a mean correction factor $\langle f(i)\rangle = 2.9$ with $f(i) = \sin i/\cos^2 i$ and $\langle i\rangle \approx 57.3\degr$. \citet{bontemps1996} applied an additional mean correction for the optical depth $\langle\tau_{\rm CO}(1-e^{-\tau_{\rm CO}})\rangle$ of 3.5 based on opacity values found by CB92, but opacity effects are not part of this comparison, so this factor is not included:
\begin{equation}
F_{\rm CO} = \langle f(i)\rangle\times \langle\frac{\tau_{\rm CO}}{1-e^{-\tau_{\rm CO}}}\rangle \times \frac{1}{\Delta{r}} \times K\left(\displaystyle\sum\limits_j \left[\int T(\varv')\varv'^2{\mathrm d}\varv'\right]_j\right) \\
\end{equation}

\subsubsection{Separation method (M7)}
In M7, the momentum and dynamical age are considered as separate parameters, with the dynamical age estimated using the maximum velocity, while the intensity-weighted velocity is a better measure for the momentum since it takes the kinematic structure into account \citep{downes2007,curtis2010}. The dynamical age is corrected for inclination by using the ratios of $t_{\rm \varv_{max}}$/True age in Table 6 of \citet{downes2007}. For these corrections, the geometrical mean for the mass density contrast $\eta = 0.1$ and $\eta = 1.0$ cases are taken for the first column in this table, and interpolated to correspond to the derived inclination angles, where $i = 90-\alpha$ (see Table \ref{corrections}):

\begin{equation}
F_{\rm CO} = c_3 \times \frac{K\left(\displaystyle\sum\limits_j \left[\int T(\varv')\varv'{\mathrm d}\varv'\right]_j\right)\varv'_{\rm max}}{{R_{\rm lobe}}}
\end{equation}

\subsubsection{Comparison of outflow force}
\label{dispersion}
\begin{figure*}
\includegraphics[scale=0.8]{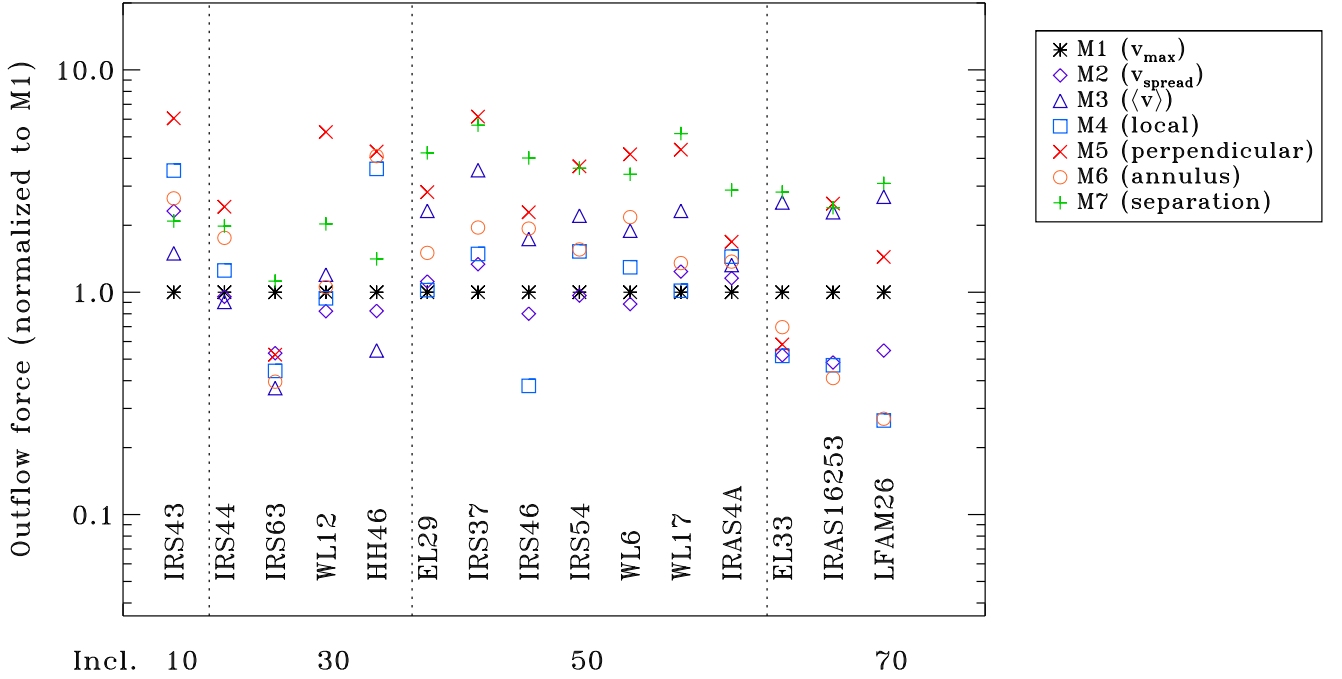}
\caption{Outflow forces derived using different methods, normalized to the $\varv_{\rm max}$ method (M1), for all Ophiuchus Class I outflows and HH~46 and IRAS~4A. The sources are sorted by inclination angle.}
\label{allmethods}
\end{figure*}

All methods were applied to the same set of observations for all detected outflows in Ophiuchus, HH~46 and the Class 0 source IRAS~4A. For IRAS~4A, only the spectra within a 50$\arcsec$ radius are integrated to exclude ``bullet'' emission \citep{yildiz2012}, see also Sect. \ref{spatialcov}.
Figure \ref{allmethods} summarizes the results, normalized to the $\varv_{\rm max}$ method (M1). Since the \emph{true} outflow force remains unknown, all results are presented compared to this method. The sources are sorted by inclination angle, which is an important variable in determining an accurate outflow force (see Sect. \ref{dispersion}).

Figure \ref{allmethods} shows the large dispersion of the resulting outflow force using different methods. All methods agree with each other within a factor of 6, with M5 and M7 giving systematically higher values than M1 with a mean factor M/M1 of 3.0, while M2, M3, M4 and M6 have mean factors of $\sim$1.5. M2 agrees very well with M1, implying that the correction factor $c_1$ taken from CB92 mainly compensates for the spread in velocity throughout the outflow lobe. The methods that do not involve correction factors (M2, M4, M5 and M6) result in lower values of the outflow force for the largest inclination angle with a factor of 2--3, but no other systematic effects related to inclination are found. IRAS~4A shows the smallest spread between methods, illustrating that the variations become less important when larger outflow velocities and smaller opening angles are involved such as usually found for Class 0 sources. 

The largest uncertainties in the derivation of the outflow force are $\varv_{\rm max}$ and inclination. The velocity $\varv_{\rm max}$ is difficult to measure precisely since the shape of the outflow wings is almost exponential. The outflow force is less affected by the choice of $\varv_{\rm max}$ for M2, due to the local $\varv_{{\rm out},j}$, whereas the weighted velocity methods are only affected if the outer velocity limit is underestimated: an overestimate of $\varv_{\rm max}$ will not result in higher values since there is little emission at higher velocities. M1 has the largest uncertainty in this case since it depends quadratically on $\varv_{\rm max}$. 

The inclination remains the biggest issue. Determining the inclination based on the morphology is fairly uncertain: generally one can only distinguish between pole-on, plane of the sky or something in between. A reasonable assumption is that $i=10\degr$ corresponds to $i\sim0-30\degr$, $i=30\degr$ to $i\sim30-50\degr$, $i=50\degr$ to $i\sim30-70\degr$ and $i=70\degr$ to $i\sim50-70\degr$, introducing intrinsic errors in M1, M3, M6 and M7. In addition, the correction factors that are based on models (M1, M3 and M7) are in themselves model-dependent, whereas outflow models are still relatively simple. M6 does not take individual inclinations into account and is therefore less accurate for the highest and lowest inclination. The correction factors from CB92 (M1, M3) are based on the spread in correction between a model of a high-velocity accelerated conical lobe surrounded by a slower envelope with three different velocity fields $\varv(r)$: $\varv(r)\propto 1/r$, $\varv(r)\propto r$ and $\varv(r)\propto {\rm constant}$. The correction factors from DC07 (M7) are based on a more advanced model: a protostellar jet model in an molecular cloud in a long-duration numerical simulation predicts the resulting outflow and its observed properties. DC07 also compares the outcome of M1 (including the corrections from CB92) and M3 (but without corrections) and they find that M1 is still correct to within a factor of 2 even though it was based on a much simpler model. M3, on the other hand, significantly underestimates the outflow force, especially for larger inclination angles, by at least a factor of 20 if no correction factor is applied while the correction factors based on the model from CB92 are only 1.2--7.1 (see Table \ref{corrections}) so these are clearly not sufficient. Surprisingly, the inclination dependence of M2 is weak, which would have been expected from the correction factors in M1.

In comparing these methods, it remains unclear which method best represents the \emph{true} physical value of the outflow force, although the Class 0 outflow models of DC07 show that M3 and M4 (named global and local method in DC07) severely underestimate the physical outflow force and any applied correction factors are largely uncertain. In their case the outcome could be compared directly to the true force, as the energetic parameters are part of the model input. On the other hand, both M1 and M7 in the DC07 (using the dynamical time) matched reasonably well with the input models, using the inclination correction factors. From these two, M7 is least affected by inclination and choice of $\varv_{\rm max}$.

\subsection{Outflow emission at ambient velocities}
\label{ambient}
\begin{figure}[ht]
\includegraphics[scale=0.4]{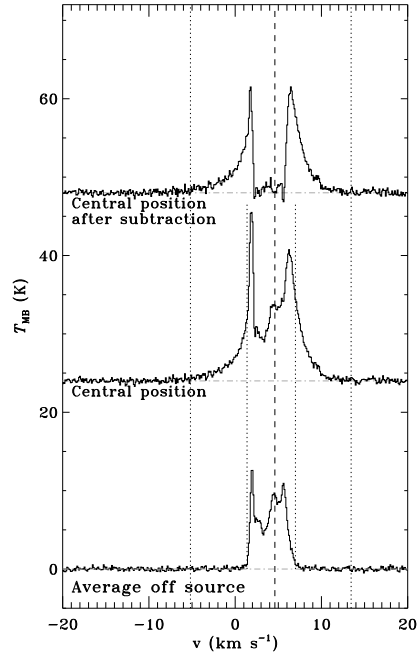}
\caption{An example of an envelope-subtracted outflow $^{12}$CO 3--2 profile toward Elias~29. From bottom to top: the average of four off-source spectra in the Elias~29 map, the spectrum of the central position and the spectrum of the central position after subtraction of the averaged off-source spectra.}
\label{subtraction}
\end{figure}

Another issue in outflow force analysis is the outflow material at ambient velocities, blending in with the low-velocity envelope material (DC07). In general, inner velocity limits are chosen at a few km s$^{-1}$ from the source velocity to exclude the bright and usually optically thick envelope emission, but in doing so some outflow material at low velocities is excluded. DC07 show that ignoring low-velocity material leads to an underestimate of the derived outflow momentum of typically a factor of 2--3 depending on the inner velocity limit and the maximum outer velocity. Some studies argue instead for subtraction of an off-source envelope spectrum and integration of the full remaining profile to obtain the outflow mass \citep[e.g.][]{bontemps1996}. The effect of subtraction of an envelope profile is investigated here (see Fig. \ref{subtraction}), making good use of our HARP maps. The envelope profile is determined by averaging a number of off-source spectra, which are selected by eye from each map. This is not trivial in some of the Ophiuchus maps, due to the confusion with nearby outflows and foreground cloud emission. Typically, very little emission is left at ambient velocities after subtraction, although this is not the case for HH~46 and IRAS~4A. The remaining profile is integrated over all non-negative emission channels up to $\varv_{{\rm out},j}$.

\begin{figure}
\includegraphics[scale=0.6]{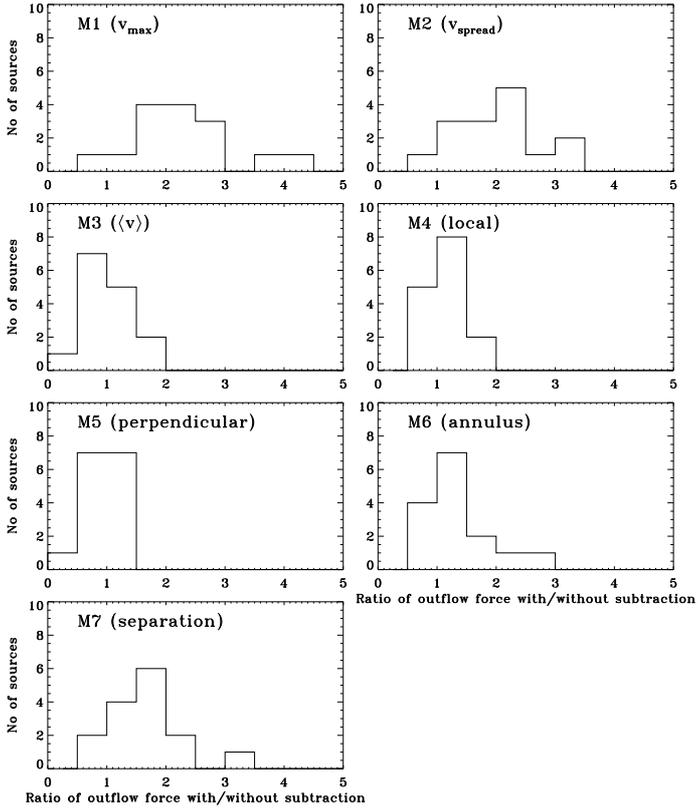}
\caption{Histogram of the ratio between the outflow force determination with subtraction of an envelope profile and without subtraction, for all sources in this study and for each method described in Section \ref{recipe}.}
\label{subtraction_hist}
\end{figure}

Figure \ref{subtraction_hist} shows the results of this comparison. Systematic differences between the methods are clearly visible: the ratio is not larger than 2 for all intensity-weighted methods (M3--M6), while the outflow force can be up to a factor of 4 higher for the $\varv_{\rm max}$ method (M1). This is an intrinsic effect of the unweighted methods: all material, including that at ambient velocities, is multiplied with the high-velocity value, while the contribution of this material is less in the intensity-weighted methods. M7 follows a distribution between M1 and M3-M6 as it is only partly an intensity-weighted method. Still, including ambient material does not increase the outflow force by more than a factor of 5, which is lower than determined from the modeling results from DC07, who find factors up to 10--100. This can be understood as follows: by integrating the full profile (after subtraction), the total outflow mass is increased by the inclusion of the ambient outflow emission. However, in the derivation without subtraction some envelope emission is still included in the integration. These two factors add a comparable amount and therefore an analysis without subtraction does not automatically underestimate the mass. 

\subsection{Other influences on the outflow force}
Apart from the analysis method, the derived outflow force can be influenced by the data quality and choice of velocity limits. The effects are studied quantitatively in Appendix \ref{influence}. Although $\varv_{\rm max}$ depends on the data quality, the outflow force does not change more than a factor of 2 for small changes in the inner velocity limit, the rms level, the outer cut-off level and spectral binning even for the M1 method, which depends mostly on $\varv_{\rm max}$. It is also demonstrated that the spatial coverage in principle does not influence the outflow force, indicating that this is a conserved quantity along the outflow direction. The determination of the parameter $\varv_{\rm max}$ can also be affected by the quality of the baseline subtraction and thus instrumental effects and/or adopted observational techniques (e.g. on-the-fly mapping). The data and the baseline subtraction presented here is shown to be of excellent quality. However, if such quality is not apparent, one should be careful in the determination of $\varv_{\rm max}$. When combining the results from different studies, using different methods, assumptions and/or data quality, scatter up to a factor of 5 can be expected.

Measuring the size $R_{\rm lobe}$ from the observed CO gas makes the incorrect assumption that the gas has travelled at a constant speed from the protostar, whereas the observed CO is actually entrained from the environment and originates from the location where it has been accelerated. As long as acceleration occurs on $<$100 AU scales, the effect should be small. Modeling the effect of the environment on the measured outflow force is beyond the scope of this work.

\subsection{Application to Ophiuchus}
The basic outflow parameters (mass, size and velocity limits) are derived for each source, for both the blue and red lobes, using the main recipe described in Sect. \ref{recipe}. The source velocities were derived using HCO$^+$ $J$=4--3  or C$^{18}$O $J$=3--2 spectra from \citet{kempen2009}. The $^{12}$CO $J$=3--2 line wing is assumed to be optically thin and no off-source spectra were subtracted. These basic parameters are not corrected for inclination.  

With the $\varv_{\rm max}$ method (M1) and the separation method (M7) the dynamical age and outflow force of all Ophiuchus sources are derived, the outflow force is corrected for inclination using the correction factors from CB92 and DC07. Although M7 is recommended for future derivation of outflow forces, the results from M1 are presented as well, as they will be combined with results from previous studies in evolutionary plots in Sect. \ref{trends}. The new outflow structures without IR source (U1, U2, etc.) are analyzed in a similar way, although the parameters have a high uncertainty due to the poor spatial coverage. For the same reason, the inclination could not be derived for these outflows. The parameters for all outflows in Ophiuchus are given in Table \ref{outflowparameters}. The outflow force for Elias~33 was derived for both the regular $2\arcmin\times2\arcmin$ jiggle map and the larger $4\arcmin\times4\arcmin$ raster map. In Table \ref{outflowparameters} only the first is given; it is within a factor of 2 with the derived value from the $4\arcmin\times4\arcmin$ raster map with $F_{\rm CO}$ = 8.3$\cdot10^{-6} M_{\odot}$ km s$^{-1}$ yr$^{-1}$, so consistent with the outflow force being a conserved quantity along the radius. 

\begin{table*}[ht]
\caption{Outflow parameters from $^{12}$CO observations for Ophiuchus sources.}
\label{outflowparameters}
\small
\begin{tabular}{lcccccccccccc}
\hline\hline
&&\multicolumn{3}{c}{Blue lobe}&&\multicolumn{3}{c}{Red lobe}&&M1&M7\\
\cline{3-5}\cline{7-9}
Name&$\varv_{\rm source}$&$\varv'_{\rm max}$&$R_{\rm lobe}$&$M$&&$\varv'_{\rm max}$&$R_{\rm lobe}$&$M$&$t_{\rm d}$\tablefootmark{a}&$F_{\rm CO}$\tablefootmark{b}&$F_{\rm CO}$\tablefootmark{b}\\
&(km s$^{-1}$)&(km s$^{-1}$)&(AU)&($M_{\odot}$)&&(km s$^{-1}$)&(AU)&($M_{\odot}$)&(yr)&($M_{\odot}$ km s$^{-1}$&($M_{\odot}$ km s$^{-1}$\\
&&&&&&&&&& yr$^{-1}$)& yr$^{-1}$)\\
\hline
WL~12&4.3&6.1&7110&6.3($-$5)&&6.8&5760&1.2($-$4)&4.8(3)&1.2($-$7)&2.5($-$7)\\
LFAM~26&4.2&5.2&3200&2.9($-$4)&&12.8&6480&2.6($-$3)&3.7(3)&2.0($-$6)&5.5($-$6)\\
WL~17&4.5&5.0&7380&2.0($-$5)&&5.3&4500&4.0($-$4)&5.5(3)&2.5($-$7)&1.3($-$6)\\
EL~29&4.6&9.8&4590&6.1($-$4)&&8.8&4590&4.0($-$4)&2.3(3)&1.9($-$6)&0.8($-$5)\\
IRS~37&4.2&6.2&5130&0.9($-$4)&&3.9&7830&4.4($-$4)&6.7(3)&1.4($-$7)&0.8($-$6)\\
WL~6&4.0&7.3&7650&2.0($-$4)&&10.3&8550&6.2($-$4)&4.5(3)&0.9($-$6)&2.9($-$6)\\
IRS~43&3.8&7.7&4950&1.4($-$4)&&7.3&5400&2.7($-$4)&3.3(3)&2.6($-$7)&5.5($-$7)\\
IRS~44&3.8&9.0&7650&4.7($-$4)&&13.7&7740&1.2($-$3)&3.4(3)&3.2($-$6)&6.4($-$6)\\
IRS~46&3.8&7.5&5130&2.3($-$4)&&10.3&4950&1.3($-$3)&2.8(3)&2.8($-$6)&1.1($-$5)\\
EL~33&4.5&11.2&8550&4.4($-$3)&&5.8&6750&1.4($-$3)&4.6(3)&1.3($-$5)&3.7($-$5)\\
IRS~54&4.1&11.2&10260&3.3($-$4)&&8.3&11520&2.5($-$4)&5.5(3)&0.9($-$6)&3.4($-$6)\\
IRAS~16253-2429&4.0&5.6&10170&1.4($-$4)&&5.8&8910&5.3($-$4)&7.9(3)&5.7($-$7)&1.4($-$6)\\
IRS~63&2.7&10&5850&4.8($-$4)&&4.6&6750&3.1($-$4)&4.9(3)&0.9($-$6)&1.0($-$6)\\
U1&2.7&4.9&5130&2($-$4)&&4.6&4500&1.7($-$4)&4.8(3)&3.7($-$7)&1.8($-$7)\\
U2&2.7&\ldots&\ldots&\ldots&&3.3&1800&5.0($-$5)&2.6(3)&6.5($-$8)&4.4($-$8)\\
U3&4.2&8.8&7740&4.9($-$4)&&4.8&7740&1.4($-$3)&5.9(3)&1.9($-$6)&0.9($-$6)\\
U4&4.3&4.6&7110&0.8($-$4)&&4.7&12960&5.4($-$4)&1.0(4)&2.4($-$7)&1.4($-$7)\\
U5&3.8&6.2&2700&0.9($-$4)&&6.0&2700&7.2($-$5)&2.1(3)&4.7($-$7)&4.4($-$7)\\
U6&3.8&6.4&3600&5.5($-$5)&&\ldots&\ldots&\ldots&2.6(3)&1.8($-$7)&1.4($-$7)\\
\hline
\end{tabular}
\tablefoot{
Numbers in brackets indicate the exponent of the power of 10 of each value.
\tablefoottext{a} Average of the two lobes.  
\tablefoottext{b} Sum of the two lobes.
}
\end{table*}

\section{Discussion}
\subsection{Outflow characterization}
Outflow activity has been associated with most of the embedded sources in this study. The only exceptions are outflows that are confused by other nearby outflows. Automatic line wing detection routines are not sufficient within clustered star formation such as Ophiuchus, especially when the envelope line profiles change within the star forming region. Earlier results of the non-detection of outflows may have been caused by the use of automatic line wing detection, low spatial resolution or low $S/N$. Since detailed inspection revealed bipolar outflows for every source in this study, it is likely that every embedded source has a bipolar outflow. This reinforces the protostellar nature of the newly detected outflow sources.

The sources for which no outflow could be identified due to confusion with another nearby source (Elias~32 and 33, IRS~37 and WL~3) are explored somewhat further. In this study, the driving source is determined based on morphology. Both sources may have outflows which are blended or along the same direction. In that case, the total outflow mass (and the outflow force) should be split up between the two sources. 

A closer look at the spectrum of Elias~33 (Fig. \ref{elias33}) shows a double structure in the line wings, especially the blue wing, with a high-intensity inner wing of 11 K and a low-intensity outer wing of 7 K, possibly indicating overlap of separate wings and suggesting blending of outflow emission of two separate sources. For IRS~37 and WL~3, the outflow force is not strong, but the envelope mass for both is $<$10 times that of Elias~33, suggesting more evolved YSOs. The outflow is however confused due to the missing data in the south east corner. Observations with full coverage will improve the possibility to assign this outflow to either one of the sources.

The bipolar outflow activity and outflow shape of all sources classified as `Stage 1' in \citet{kempen2009} do not contradict their classification methods. Outflows from more evolved Stage 2 sources are in general less bipolar and more wide-angled \citep{arce2006} than Stage 1, but the opening angle cannot be derived from our CO maps due to insufficient spatial resolution and low inclination. The difference between Stage 1 and Stage 2 can therefore not be confirmed by an outflow study, since Stage 2 sources may also show outflow activity, for example IRS~46 (in this study), IRS~48 and IRS~51 \citep{bontemps1996} and WL~10 \citep{sekimoto1997}. No statistics are known for the occurrence of outflows with Stage 2 sources, but a non-detection would have been surprising for a Stage 1 source since all embedded protostars are expected to go through an outflow phase \citep{fukui1993,arce2006}. 

When comparing the outflow forces derived in this paper with values in previous studies, significant differences are found, up to almost two orders of magnitude (see App. \ref{compareof}). These differences can be explained to within a factor of a few by a detailed look into the methods and assumptions used previously. Since uncertainties in opacity and inclination potentially cause the largest variations in the outflow force, it is essential to derive these properties as accurately as possible. 

Since the jet driving the outflow is assumed to originate from the spinning up of the magnetic field by the rotating disk, the outflow direction is perpendicular to the disk \citep{matsumoto2004}. Therefore, the morphology of outflow and disk observations can be directly compared. In the last decade, very high spatial resolution observations revealing the disk structure of early phase YSOs have become available \citep{brinch2007,lommen2008,jorgensen2009} to which we compare the outflow results of this study. 

The outflow of IRS~63 matches well with a disk inclination of 30$\degr$ and north-south orientation as was found with interferometric HCO$^+$ $J$=3--2 observations \citep{lommen2008,jorgensen2009}. The outflow of IRS~46, which is in the plane of the sky, is also consistent with an edge-on disk, as was determined by fitting a disk model to the SED \citep{lahuis2006}. In contrast, the outflow of Elias~29 is inconsistent with the interpretation of the HCO$^+$ velocity gradient of \citet{lommen2008} and \citet{jorgensen2009}, who argued that the blue-shifted and red-shifted emission are indicative of a disk ($i=30\degr$) in the north-north-west, south-south-east direction \citep[see][Fig. 5]{jorgensen2009}. As this direction is parallel to the outflow emission, it is more likely that HCO$^+$ is instead tracing the dense swept-up outflow material, indicating a small jet very close to the source. IRS~43 was interpreted as a nearly edge-on disk ($i\sim$90$\degr$) in west-north-west, east-south-east direction \citep{jorgensen2009}, further supported by the elongated structure of continuum, HCO$^+$ and HCN emission, the direction of nearby Herbig Haro objects \citep{grosso2001} and the proposed radio thermal jet \citep{girart2000}. However, this does not fit at all with the outflow result of a nearly pole-on outflow ($i\sim$10$\degr$); the outflow with an edge-on disk would be highly inclined and therefore not even visible because the projected velocity is too small. As both the disk material and the swept-up outflow material are clearly observed, the configuration may be more complex. It is possible that the disk and surrounding envelope are misaligned, similar to the L1489 IRS system \citep{brinch2007}, as IRS~43 is a close near infrared binary with a separation of 0.6$\arcsec$ \citep{duchene2007,herczeg2011}. In that case,  the HCO$^+$ emission would trace the inner circumbinary envelope, but not the actual disk. The Herbig Haro objects that are associated with IRS~43 by \citet{grosso2001} are also consistent with the outflow direction of IRS~44, to the north of IRS~43 \citep[see][Fig. B1]{jorgensen2009}.

\subsection{Trends and evolution}
\label{trends}
\begin{figure*}[tp]
\includegraphics[width=18cm]{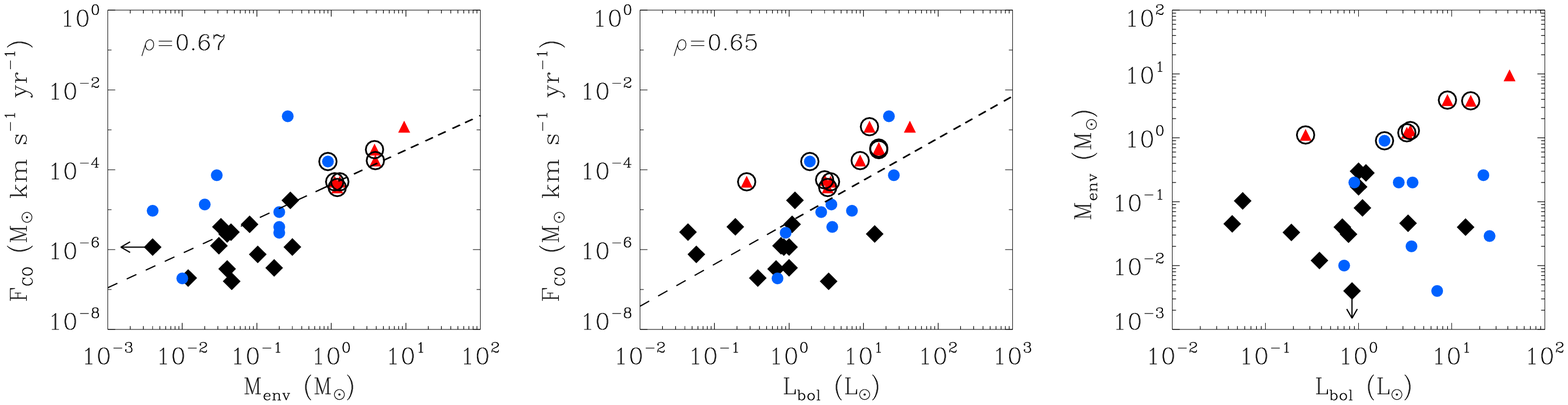}
\caption{Correlation plots for the outflow force $F_{\rm CO}$ using M1, bolometric luminosity $L_{\rm bol}$ and the mass of the envelope $M_{\rm env}$. Black diamonds are from this study, red triangles from CB92, and blue circles from \citet{hogerheijde1998}, where $F_{\rm CO}$ was calculated by the same method as in this study. The Class 0 sources are encircled. Upper limits are indicated with arrows. The dashed line indicates the best fits to the combined data set.}
\label{vsplots} 
\end{figure*} 
In order to gain a better understanding of the outflow driving mechanism, two main trends are explored: the outflow force versus the mass of the envelope and the outflow force versus the bolometric luminosity, using the outflow force results from the $\varv_{\rm max}$ (M1) method. The envelope mass is considered an evolutionary parameter, whereas the bolometric luminosity for Class 0/I YSOs is dominated by accretion luminosity \citep{bontemps1996}. 

The correlation plots are shown in Fig. \ref{vsplots}, where data from this study are combined with data from two other outflow studies, which used the same method (M1) for deriving the outflow force, i.e., CB92 (sample of mostly Class 0 sources) and \citet{hogerheijde1998} (mostly Class I sources in Taurus). From the CB92 sample we have only included the low mass sources. T Tau and L1551-IRS5 appear in both CB92 and \citet{hogerheijde1998}, we have chosen to include only the latter (the values for $F_{\rm CO}$ are within a factor of 2 of each other). HH7-11 in CB92 was later classified as Class I, while L1527 IRS in \citet{hogerheijde1998} was classified as a Class 0 source. These sources are marked accordingly in the plot. \citet{hogerheijde1998} and CB92 measured the opacity of the line wings from $^{13}$CO observations and corrected for this when determining the outflow mass. 

The well-known relationship between envelope mass and outflow force \citep{cabrit1992,bontemps1996,hogerheijde1998,hatchell2007} is further extended with the results of this study. Envelope masses for our sample (see Table \ref{sample})
were taken from \citet{kristensen2012} where available, otherwise from \citet{kempen2009}. The values from \citet{kristensen2012} are a better estimate due to full modeling of the SED using DUSTY, whereas the envelope masses from \citet{kempen2009} were determined using a conversion with a single 850 $\mu$m flux. Envelope masses and classification for the CB92 and Hogerheijde sample were updated with more recent values, see Table \ref{trendparameters}. The best fit through all data points is $\log(F_{\rm CO}) = (-4.4\pm0.2) + (0.86\pm0.19)\log(M_{\rm env})$ with Pearson's correlation coefficient $\rho_{F_{\rm CO},M_{\rm env}}$=0.67. The relationship agrees within errors with the large range trend found by \citet{bontemps1996}. The data points from this study are offset from the other two studies, which may be due to the data used for deriving the envelope mass: the envelope masses in this study, except for Elias~29 and IRS~63, were derived from SCUBA 850 $\mu$m emission \citep{kempen2009}, while \citet{hogerheijde1998} based their envelope mass on 1.3 mm emission. The correlation coefficient is the same as was found by \citet{bontemps1996}, who uses the annulus method (M6) and a constant inclination correction, confirming our previous statement that the choice of method introduces small scatter in the values for the outflow force, but no significant changes over a large range. The decline of outflow force with envelope mass may reflect a decrease of the outflow force with evolution \citep{bontemps1996,saraceno1996}, but since the current envelope mass reflects both age and initial core mass, the range of envelope masses in this sample could also represent different initial conditions \citep{hatchell2007}, so the link with evolution is less obvious. 

A second well-known relationship is between outflow force and bolometric luminosity, given in the middle panel of Fig. \ref{vsplots}. Bolometric luminosities for our sample (see Table \ref{sample})
were taken from \citet{kristensen2012} where available, otherwise from \citet{kempen2009}. The values from Kristensen et al.\ are better constrained due to the inclusion of \emph{Herschel} PACS far-infrared data. The best fit through the combined data points is $\log(F_{\rm CO}) = (-5.3\pm0.2) + (1.1\pm0.2)\log(L_{\rm bol})$ with $\rho_{F_{\rm CO},L_{\rm bol}}$=0.65.  
The correlation with luminosity is usually interpreted as evidence that the driving mechanism for molecular outflows is directly related to the accretion process, since the bolometric luminosity of low-mass YSOs is thought to be dominated by accretion luminosity \citep{lada1985,yorke1995}. It is widely thought that outflows are momentum-driven by a jet or wind originating from the inner disk or protostar \citep{bontemps1996,downes2007}. The most plausible energy source for this jet or wind is the gravitational energy released by accretion onto the protostar \citep{bontemps1996}. Accordingly, the outflow force is directly related to the accretion rate $\dot{M}_{\rm acc}$. 
\citet{bontemps1996} found a decline of outflow force between Class 0 and Class I, which has been taken as evidence that outflow force declines with age. In our combined sample the Class 0 sources are also offset from the Class I sources and fitting these two classes separately results in a significant change in offset ($-$5.6 vs $-$4.2, $\pm$0.2) for Class I vs 0). However, due to the selection of the older samples (the brightest outflow sources available) and the differences in data quality and derivation of parameters, the combined sample is probably not representative of the general properties of Class 0 and Class I sources.

The envelope mass as function of bolometric luminosity has been considered as an evolutionary diagram \citep{saraceno1996} since the parameters are correlated, but \citet{bontemps1996} concluded that Class 0 sources do not follow the correlation and the diagram can only show the range of evolutionary stages of the sample. For our sample these two parameters only have a $\rho_{M_{\rm env},L_{\rm bol}}$ of 0.34 in combination with other studies (see right panel of Fig. \ref{vsplots}) so their correlations with $F_{\rm CO}$ are independent. The similar correlation coefficients for these two relations have 3$\sigma$ confidentiality levels considering the sample size \citep{marseille2010} so we consider the correlations to be strong.

In summary, the correlation plots agree well with previous studies of the outflow force, but a decline in evolution between Class 0 and I objects is neither confirmed or ruled out.

\subsection{Outflow direction}
For the L1688 part of Ophiuchus, we compare the outflow directions for the different sources using Fig. \ref{bigmap}. Since these sources are clustered together, they may have experienced the same trigger from the same direction for the core or filament formation and the following star formation. For L1688, the orientation can be divided in three groups: IRAS~16253--2429, IRS~54, IRS~37, WL~6, IRS~44 and IRS~46 are all oriented in a north-east, south-west direction. We may even add U3 (near IRS~37) and U5 (near IRS~44) to this sample, which are well enough covered to derive their orientation. The second group contains the sources with a north-west, south-east direction: Elias~29, Elias~33, VLA~1623, WL~12 and WL~17. LFAM~26 and IRS~43 do not belong to either group. The direction of the first group agrees with the results of \citet{anathpindika2008}, who concluded that the outflow direction is perpendicular to the filament direction in 72$\%$ of cases. The implication is that angular momentum is delivered to a core forming in a filament, since the angular momentum will eventually drive the outflow. The Ophiuchus ridge in the south is indeed perpendicular to the outflow direction of the first group. A detailed study of filament velocities, such as performed for L1517 \citep{hacar2011} suggesting core formation in two steps, could provide a better understanding of this phenomenon.

The distribution of outflows over two groups with an approximately equal direction suggests that there have been two separate triggering events causing the star formation or two separate filaments. Considering the large values for $L_{\rm bol}$ and $M_{\rm env}$ for the second, northern group compared to the first, southern group (mean of 3.9 $L_{\odot}$ and 0.36 $M_{\odot}$ versus 0.6 $L_{\odot}$ and 0.044 $M_{\odot}$) these two groups may have started star formation at different times, due to different events. This is consistent with the observation of \citet{zhang2009} that the star formation in Ophiuchus first took place in the denser northwestern L1689 region. It is further consistent with the suggestion that star formation in Ophiuchus was triggered by ionization fronts and winds from the Upper Scorpius OB association, located to the west of the Ophiuchus cloud \citep{blaauw1991,preibisch1999,nutter2006}. 

\section{Conclusions}
We have searched for molecular outflows towards Class I sources in Ophiuchus using new high $S/N$ JCMT $^{12}$CO $J$=3--2 maps and compared various analysis methods for the outflow force and assumptions that go into the calculation. The main results of this study are the following:
\begin{enumerate}
\item All embedded sources classified as `Stage 1' by \citet{kempen2009} show bipolar outflow activity, except for those that are so close to another source that their outflows are confused. Five new outflows are detected. These results are consistent with the fact that every embedded source likely has a bipolar outflow. The methods used for determining integration limits and deriving physical properties strongly influence the results of outflow studies, but for large data sets over broad ranges of $L_{\rm bol}$ and $M_{\rm env}$ the trends are still similar. For weak outflows and clustered star formation, detailed analysis and high resolution observations are crucial.
\item Seven different analysis methods for deriving the outflow force are analytically described and applied to all Ophiuchus sources in the sample, plus the well-studied sources HH~46 and NGC1333 IRAS~4A. All methods agree with each other within a factor 6. The methods that do not involve inclination correction factors give lower values for the outflow force for the largest inclination angle by a factor 2--3, but no other systematic effects related to inclination exist. Although the true outflow force remains unknown, the separation method (separate calculation of dynamical time and momentum) is least affected by the uncertain observational parameters.
\item The effect of subtraction of an off-source spectrum, representative of an envelope profile, is studied. After this subtraction, the outflow material at ambient velocities, blended in with the envelope material can be measured. It is found that including ambient material does not increase the outflow force by more than a factor of 5 and generally much less.
\item The outflow force remains constant as a function of radius, so the outflow force can be analyzed with partial coverage CO maps as long as it is centered on the source position.
\item Observational properties and choices in the analysis procedure can individually affect the outflow force up to a factor of a few. The most important factor to consider is the $S/N$ of the data: a strong dependence of the outflow force on the noise level means that the methods described cannot always be applied directly. 
\item When combining the results from different studies, using different methods, assumptions and/or data quality, scatter up to a factor of 5 can be expected. Discrepancies in derived outflow force between different studies are up to an order of magnitude and can be explained by comparing the analysis methods.
\item Comparing the outflow observations for three sources with recently obtained disk studies \citep{jorgensen2009} leads to revision of these disk structures. IRS~63 shows an excellent agreement for the disk orientation (perpendicular to the outflow direction), but for Elias~29 the assumed disk emission most likely originates from the outflow material itself, close to the source, consistent with a non-detection in millimetre continuum at long baselines. IRS~43 is even more complex: the outflow direction was found to be pole-on, which is completely inconsistent with the previously identified edge-on disk orientation.
\item The well-known correlations of outflow force with envelope mass and bolometric luminosity are extended with the results of this study, confirming a direct correlation of outflow strength with both properties.
\item The outflows in the L1688 region can be divided into two groups, based on preferred outflow direction and significantly different evolutionary properties. This suggests a scenario with star formation in two separately triggered events, starting in the north west, supporting conclusions from previous work, e.g. \citet{zhang2009}.
\end{enumerate}

For further outflow studies, it is important to consider the choice of methods and assumptions that go into the calculations, especially when comparing with previous results. For Ophiuchus, a complete map of the entire L1688 region in low-$J$ CO lines, as recently performed for Perseus \citep{curtis2010}, would provide a more complete view of outflow activity, both for known as well as for new sources (such as the Us) and a more complete view of outflows from Stage 2 sources. Furthermore, the envelope surrounding IRS~43 should be explored in very high spatial resolution in order to understand the complex velocity structure. Limitations by single dish observations will be solved further using the Atacama Large Millimeter Array (ALMA). 
\begin{acknowledgements}
The authors would like to thank Sylvie Cabrit and Mario Tafalla for useful discussions, Antonio Chrysostomou who carried out part of the $^{12}$CO observations and Rowin Meijerink and Edo Loenen who carried out the $^{13}$CO observations. The James Clerk Maxwell Telescope is operated by the Joint Astronomy Centre on behalf of the Science and Technology Facilities Council of the United Kingdom, the Netherlands Organisation for Scientific Research, and the National Research Council of Canada. Astrochemistry in Leiden is supported by the Netherlands
Research School for Astronomy (NOVA), by a Spinoza grant and grant 614.001.008 from the Netherlands Organisation for Scientific Research (NWO).
\end{acknowledgements}

\bibliographystyle{aa}
\bibliography{myrefs}

\appendix

\section{Newly discovered outflows}
\label{Us}
From detailed spectral analysis, new bipolar outflow structures were discovered in several CO maps, not belonging to any of the sources in the embedded source sample of \citet{kempen2009}. In this section we look for other suitable candidates of embedded sources.  The partial coverage of these outflows extends the field in which to look for a candidate. 
\begin{itemize}
\item The outflow structure U1 south west of IRS~63 may originate from the submillimeter core SMM J 163133--24032, which was matched by \citet{jorgensen2008} with the \emph{Spitzer} YSO candidate SSTc2d J163134.29-240325.2 \citep{evans2009}, located 24$\arcsec$ away, classified as a Class I source with $L_{\rm bol}$ = 0.88 $L_{\odot}$ and $T_{\rm bol}$ = 870 K.

\item For U2, selecting a candidate is somewhat difficult, as only a small piece of the red lobe is covered and therefore the direction of the originating source is unknown. The most likely candidate is the submillimeter core SMM J 163143--24003 \citep{jorgensen2008} at 2.2$\arcmin$ north east of IRS~63 (see Fig. \ref{figUs}) with a dust temperature of 15 K. No infrared source is known near this location. 

\item U3 is located to the west of IRS~37. The elongated structure and direction of the outflow suggest that the powering source is located to the south west. The large submillimeter core SMM J162713--24295 \citep{jorgensen2008} is a likely candidate (see Fig. \ref{figUs}) with a dust temperature of 18 K. The shape of this submillimeter core is consistent with the outflow direction, which is usually perpendicular to the ridge \citep{anathpindika2008}. Again, no infrared source is known at this location. 

\begin{figure*}[tp]
\includegraphics[width=16cm]{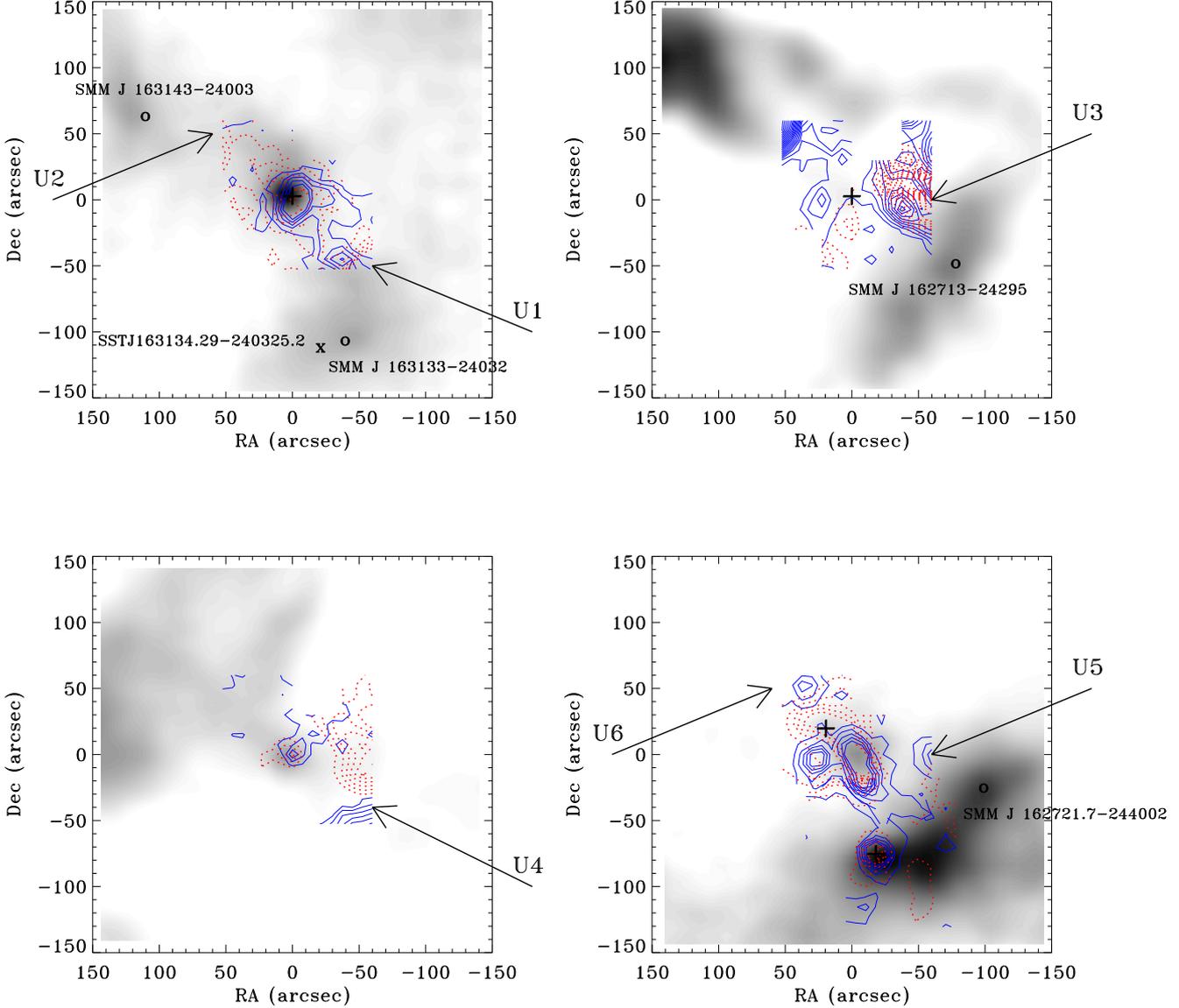}
\caption{The regions around the new outflow structures (Us), showing the SCUBA 850 $\mu$m emission in the background, the contour map of the integrated line wings, the sources from the sample in this study marked with pluses and the Us indicated with arrows. The nearby submillimeter cores are marked with circles and infrared sources with a cross. (Top left) The region around IRS~63. (Top right) The region around IRS~37. (Bottom left) The region around WL~12. (Bottom right) The region around IRS~44.}
\label{figUs}
\end{figure*}

\item The outflow U4 is clearly visible in the WL~12 map (Fig. \ref{figUs}). The center of origin is about 16:26:40, -24:35:27. There are no submillimeter cores or infrared sources nearby. Estimates of a SED or $T_{\rm bol}$ are therefore not possible at this time.

\item U5 is located to the west of IRS~44 (Fig. \ref{figUs}). Towards the south-west is submillimeter core J162721.7-244002, \citep{difrancesco2008}, which is a possible candidate for the powering source. However, just like U3, there is no infrared source at this position. No temperature was derived for this core, but the 450 and 850 $\mu$m fluxes in combination with the non-detection at 70 $\mu$m and shorter wavelengths indicate a very low dust temperature between 5 and 10 K. 

\item Only the blue lobe of U6 is detected, but there are no submillimeter cores or infrared sources nearby. Estimates of a SED or $T_{\rm bol}$ are therefore not possible at this time.

\end{itemize}

The Us are clearly peculiar objects. Only U1 could be assigned to a YSO candidate, a submillimeter core with an infrared detection. U2, U3 and U5 are interesting since they have a possible source of origin in submillimeter cores without infrared detection, suggesting either a \emph{very} low temperature and luminosity, a very deeply embedded object, an edge on geometry or a new kind of object, producing an outflow or some other mechanism producing swept-up gas. For U4 and U6 not even a submillimeter core was detected. U2, U5 and U6 are only marginally covered and therefore possibly no new outflows: they may be part of the main outflow in the map and only seem separate due to density variations in the envelope. However, for U1, U3 and U4 this is not very likely considering their morphology. Full spatial coverage of low-$J$ CO lines of these regions may provide an answer. Far-infrared fluxes provided by \emph{Herschel} may help to define the cold cores \citep[e.g.]{bontemps2010} and derive their properties, to conclude whether they can be the powering sources of these outflows. 

\newpage

\newpage

\section{Comparison with other outflow studies}
\label{compareof}
\begin{center}
\begin{table*}[tp]
\caption{Comparison of the outflow force between this study and other outflow studies.}
\label{comparestudies}
\centering
\begin{tabular}{lrcc}
\hline
\hline
Source&$F_{\rm CO}$\tablefootmark{a}&$F_{\rm CO}$\tablefootmark{b}&References\\
&\multicolumn{2}{c}{(10$^{-6}M_{\odot}$ km s$^{-1}$ yr$^{-1}$)}\\
\hline\\
WL~12&0.12&6.0&1\\
LFAM~26&2.0\phantom{5}&16,4.7&2,3\\
WL~17&0.25&\ldots&\ldots\\
Elias~29&1.9\phantom{5}&15,9.8,14.5,2.0&1,2,3,4\tablefootmark{c}\\
IRS~37&0.14&\ldots&\ldots\\
WL~6&0.9\phantom{5}&$<$15,2.5&1,4\\
IRS~43&0.26&15&1\\
IRS~44&3.2\phantom{5}&27,3.4&1,4\\
IRS~46&2.8\phantom{5}&\ldots&\ldots\\
Elias~32,33&13\phantom{.44}&9.6,146&3,5\\
IRS~54&0.9\phantom{5}&\ldots&\ldots\\
IRAS~16253--2429&0.57&1\\
IRS~63&0.9\phantom{5}&6.3\tablefootmark{d}&1\\
\hline\\
\end{tabular}
\tablefoot{
\tablefoottext{a} This study (M1)
\tablefoottext{b} Other studies (see last column)
\tablefoottext{c} The outflow force in \citet{nakamura2011} is taken from the CO 3--2 observations, by summing the blue 
and red lobes in their Table 2.
\tablefoottext{d} IRS~63 is named L1709B (name of the core) in \citet{bontemps1996}.
}
\tablebib{ 
(1) \citet{bontemps1996}, (2) \citet{bussmann2007}, (3) \citet{nakamura2011}, (4) \citet{sekimoto1997}, (5) \citet{kamazaki2003}
}
\end{table*}
\end{center}
\begin{center}
\begin{table*}[tp]
\caption{Comparison of the methods and observation properties between this study and other outflow studies.}
\label{comparepropstudies}
\centering
\small
\begin{tabular}{lccc}
\hline
\hline\\
&This study&\citet{bontemps1996}&\citet{sekimoto1997}\\
\hline
$^{12}$CO line&3--2&2--1&2--1; 1--0\\
Sources\tablefootmark{a}&All&EL~29, IRS~43, IRS~44, WL~6, WL~12&EL~29, IRS~44, WL~6\\
Beam size ($\arcsec$)&15&30(NRAO), 10(IRAM)&34\\
Map size ($\arcsec\times\arcsec$)&120$\times$120&60x60(NRAO), 25x25(IRAM)&68$\times$68\\
Velocity res. (km/s)&0.1&0.65(NRAO), 0.26(IRAM)&0.06; 0.05\\
$\sigma_{\rm rms}$ (K)&0.15&0.25(NRAO), 0.15(IRAM)&0.2; 0.15\\
Method&$\varv_{\rm max}$ (M1)&annulus (M6)&$\varv_{\rm max}$ (M1)\\
$\varv_{\rm in}$&off-source&$T_{\rm peak}/10$&unknown\\
$\varv_{\rm out}$&1$\sigma$ cut-off&1$\sigma$ cut-off&unknown\\
Opacity corr.&none&mean of CB92: 3.5&derived from $^{13}$CO: $\sim$3.0(2--1)\\
Inclination corr.&factors CB92&average $i$=57.3: factor 2.9&factors CB92\\
Distance to Oph (pc)&120&160&160\\
Other&&Average off-source spectrum subtracted&\\
&&before integration, integration &\\
&&radius 45$\arcsec$(IRS~44), 15$\arcsec$(others)&\\
\hline\\
\end{tabular}
\begin{tabular}{lccc}
\hline
\hline\\
&\citet{kamazaki2003}&\citet{bussmann2007}&\citet{nakamura2011}\\
\hline
$^{12}$CO line&3--2&3--2&3--2\\
Sources$^{\rm a}$&Elias 32/33&EL~29, LFAM~26&Elias~29,Elias~32/33\tablefootmark{b},LFAM~26\\
Beam size ($\arcsec$)&14&11&40\\
Map size ($\arcsec\times\arcsec$)&$120\times240$ (OphB2)&$300\times300$&$23\arcmin\times 23\arcmin$\\
Velocity res. (km/s)&0.33&0.4&0.4\\
$\sigma_{\rm rms}$ (K)&0.79 (OphB2)&0.4&0.28\\
Method&$\langle\varv\rangle$ (M3)&$\varv_{\rm max}$ (M1)&$\varv_{\rm max}$ (M1)\\
$\varv_{\rm in}$&channel maps\tablefootmark{c}&unknown&channel maps\\
$\varv_{\rm out}$&3$\sigma$ cut-off&unknown&3$\sigma$ cut-off\\
Opacity corr.&5.4\tablefootmark{d}&none&none\\
Inclination corr.&average $i$=57.3: factor 2.9&none&$\frac{\sin i}{\cos i}$, $i$=80 (El~29, LFAM~26), $i$=57.3 (El~32)\\
Distance to Oph (pc)&160&120&125\\
Other&&&\\
\hline\\
\end{tabular}
\tablefoot{
\tablefoottext{a} Only the sources relevant for this study are listed in this table.
\tablefoottext{b} Elias~32 was adopted by \citet{nakamura2011} as the driving source for this outflow, where Elias~33 was adopted in this study.
\tablefoottext{c} By determining integration limits using channel maps, the outflow emission is defined as the emission showing bipolar structure in a channel map. This is equivalent to the method used in this study, comparing off-source spectra with outflow spectra.
\tablefoottext{d} The factor 5.4 is based on the mean factor 3.5 as used in \citet{bontemps1996}, multiplied by 1.6 because the optical depth of the 3--2 transition is 1.6 times larger than the 2--1 transition of the latter \citep{kamazaki2003}.
}
\end{table*}
\end{center}

Of our sample of 16 sources, 11 sources were studied previously \citep{bontemps1996,sekimoto1997,kamazaki2003,bussmann2007,nakamura2011} but often with different derived properties. We compare the results, methods and available observations in Table \ref{comparestudies} and \ref{comparepropstudies}. 

The sources studied by \citet{bontemps1996} have a systematic offset in $F_{\rm CO}$ due to their opacity correction of 3.5. If this factor is removed, their outflow forces are still typically a factor of 2 higher if we use the same annulus method (M6) on our data set (see Fig. \ref{allmethods}). Since their rms level is similar to our study and other effects due to coverage or method can be excluded, the only other explanation may be the determination of the inner velocity limit: Bontemps et al.\ used $T_{\rm peak}/10$ for $\varv_{\rm in}$, but $T_{\rm peak}$ is source dependent and easily affected by the presence of foreground clouds. When the inner velocity limit is moved 1 km s$^{-1}$ inwards, the outflow force is increased by a factor of 2 (see Fig. \ref{allchoices}). The inclination factor of 2.9 for a constant $i=57\degr$ does not cause significant underestimates, since none of these five sources have an inclination of 70$\degr$ which generally requires a larger correction factor (see Fig. \ref{allmethods}).

The values found by \citet{sekimoto1997} agree very well (within a factor of 1.8) with the results in this study, as expected considering their methods and data quality. However, they derived an opacity in the line wings from $^{13}$CO lines of 3.0 for the 2--1 lines, similar to CB92.
The results for LFAM~26 and Elias~29 in the study from \citet{bussmann2007} are both a factor of $\sim$5 higher. Inspection of the individual parameters shows that this difference is dominated by the effect of integrated mass, which is more than 10 times larger in their study, whereas their dynamical time is only twice as long compared to our results. The observations presented by Bussmann et al.\ cover the entire outflow of Elias~29 and LFAM~26. Apparently the mass is not uniformly spread over the outflow lobes, so the outflow momentum may not be conserved along the lobe in this case. 

The outflow force for Elias~33 also agrees well with the value found by \citet{kamazaki2003}, considering their opacity correction of 5.4. Considering their high rms level and 3$\sigma$ cut-off, their derived outflow velocities are only 5.6 and 3.6 km s$^{-1}$, for the blue and red lobe, respectively, versus 12.0 and 10.0 km s$^{-1}$ in this study. This would decrease the outflow force with a factor of 4--5, but this is compensated by the total (uncorrected) mass derived by Kamazaki et al.\, which is more than 5 times larger than derived in this study. 

Our results do not match well with the results from \citet{nakamura2011}: LFAM~26 is comparable, but our value for Elias~29 is 7 times smaller while the value for Elias~32/33 is 2 times larger. The reason for this discrepancy may be the choice of inclination angles: the adopted value of 80$\degr$ for Elias~29 and LFAM~26 with the projection formula ${\sin i}/{\cos i}$ used by Nakamura results in a correction factor of 6, the adopted value of 57$\degr$ for Elias~32/33 gives a correction of 1.6, while in this study an inclination of $50\degr$ was assumed for Elias~29 (correction factor 0.45) and $i=70\degr$ for Elias~32/33 (correction factor 1.6). The 3$\sigma$ cutoff and low $S/N$ data used by Nakamura et al.\ would have resulted in lower values of the outflow force for all three sources with a factor 2--4, but this is counterbalanced by the use of larger correction factors (6 versus 1.1 and 1.6 versus 0.45).

This section shows again that, in order to compare and combine outflow studies, similar methods have to be applied, since results from the studies discussed can differ by more than one order of magnitude. Since opacity and inclination cause the largest variations in the outflow force, it is essential to derive these properties as accurately as possible.

\section{Mass derivation}
\label{derivemass}
The main physical parameters of the outflow are the mass, $M$, the velocity, $\varv_{\rm CO}$ (defined as $|\varv_{\rm out}-\varv_{\rm source}|$), and the projected size of the lobe $R_{\rm lobe}$, for both the blue and red lobe. The mass is calculated from the column density assuming an excitation temperature of 100 K \citep[e.g.][]{yildiz2012}. The column density of the $^{12}$CO upper level $u$  (in this case $J$=3) is derived from the integrated intensity of the line wings as follows.
\begin{equation}
N_{\rm u} = \frac{8\pi k\nu^2 \int T_{\rm mb} (\varv) \mathrm{d}\varv}{hc^3A_{\rm ul}} 
\end{equation}
The column density of the upper level, $N_{\rm u}$ (cm$^{-2}$), is converted to ${\rm H}_2$ column density $N$ by 
\begin{equation}
N = N_{\rm u} \times \left[\frac{{\rm H}_2}{^{12}{\rm CO}}\right] Q(T) [g_{\rm u} e^{-E_{\rm u}/kT}]^{-1} \\
\end{equation}
with $Q(T)$ the partition function and using ${\rm H}_2/^{12}{\rm CO}=1.2\times10^{4}$ \citep{frerking1982} and $N$ is converted to mass in grams using
\begin{equation}
M = \mu m_H A\times N
\end{equation}
with a mean molecular weight $\mu$ of 2.8 used to include He \citep{kauffmann2008} and $A$ the physical area covered by one pixel. Summarizing, the conversion from integrated intensity to mass is a multiplication with constant $K$:
\begin{equation}
K = \mu m_H A\times \left[\frac{{\rm H}_2}{^{12}{\rm CO}}\right] Q(T) [g_{\rm u} e^{-E_{\rm u}/kT}]^{-1} \frac{8\pi k\nu^2}{hc^3A_{\rm ul}} 
\label{convmass}
\end{equation}

With these basic physical quantities, a number of outflow parameters can be derived: the momentum, $p$, the dynamical age, $t_{\rm d}$, the mass outflow rate, $\dot{M}$ and the outflow force, $F_{\rm CO}$. The basic equations are:
\begin{eqnarray}
p &=& M\varv_{\rm CO} \\
t_{\rm d} &=& \frac{R_{\rm lobe}}{\varv_{\rm CO}} \\
\dot{M} &=& \frac{M}{t_{\rm d}} \\
F_{\rm CO} &=& \frac{M\varv_{\rm CO}^2}{R_{\rm lobe}} = \frac{p}{t_{\rm d}} \\
\end{eqnarray}

The velocity, $\varv_{\rm CO}$, often called $\varv_{\rm max}$, is not a well-constrained parameter: due to the inclination and the shape of the outflow, often following a bow shock including forward and transverse motion, the measured velocities along the line of sight are not necessarily representative of the real velocities driving the outflow \citep{cabrit1992}. This is the main reason why various methods have been developed to derive the outflow force, where a correction factor is often applied to compensate for these effects.

\section{Influence of assumptions and data quality}
\label{influence}
\begin{figure}
\includegraphics[scale=0.5]{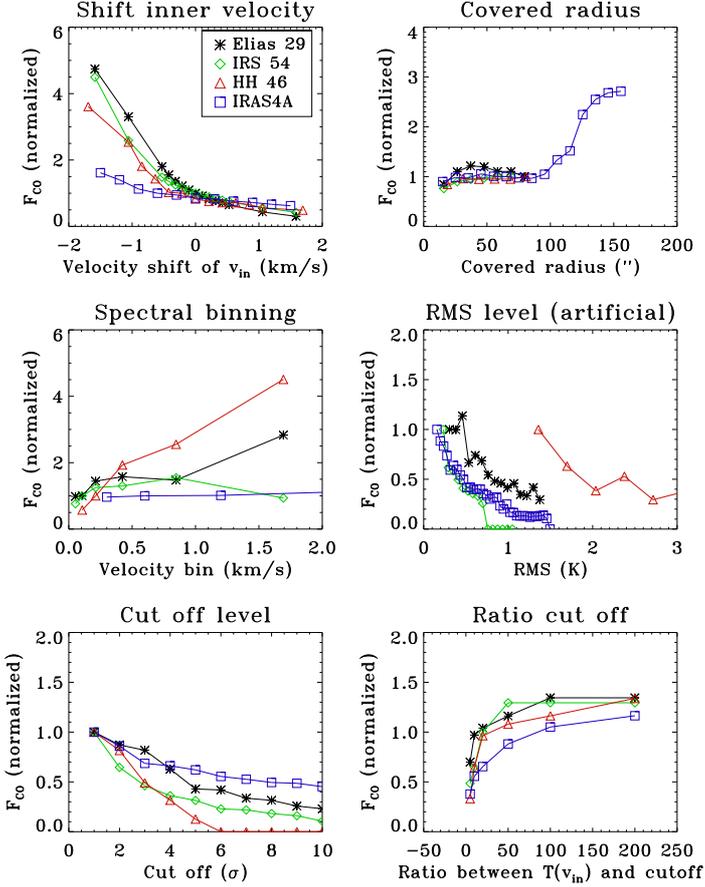}
\caption{Influence of observational properties and choices on the resulting value of the outflow force, using M1 ($\varv_{\rm max}$) on Elias~29, IRS~54, HH~46 and IRAS~4A, which all have clear, isolated outflows. See text for description of each category. It follows that these observational properties and choices in the analysis procedure affect the outflow force up to a factor of 4--5.}
\label{allchoices}
\end{figure}

Apart from the analysis method, a number of choices, observational properties and assumptions may influence the value of the outflow force. The effects are tested using the $\varv_{\rm max}$ method (M1) on a small sample of four targets: Elias~29, IRS~54, HH~46 and IRAS~4A. The results are presented in Fig. \ref{allchoices}. In each plot, the resulting outflow force is normalized to the `standard' value, which is the outflow force resulting from the best estimate of that parameter using the recipe in Sect. \ref{recipe}.

\subsubsection{Inner velocity limits}
The inner velocity limits are determined from the width of the (averaged) off-source spectra. In general, it can be difficult to determine where the outflow emission blends in with the ambient envelope emission (see also Sect. \ref{ambient}). The first panel shows how the outflow force changes if the inner velocity limits are shifted simultaneously inwards (negative) or outwards (positive), relative to the best guess based on off-source emission. When the inner limits are moved inwards, more and more integrated emission is taken into account, also including more envelope emission, increasing the mass of the outflow up to a factor of 5, but certainly within 0.5 km s$^{-1}$ the significance remains within a factor of 2. Moving the limits outwards decreases the outflow force by a factor of 2. For IRAS~4A, the outflow force increases only slightly moving inwards: due to the high velocities involved, the integrated mass is not the dominant parameter. For the intensity-weighted methods the change of the inner velocity limit changes the outflow force even less since the intensity at low velocities does not contribute much to the total outflow force.

\subsubsection{Spatial coverage}
\label{spatialcov}
\citet{bontemps1996} suggested that it does not matter how much of the outflow emission is spatially covered for calculating the outflow force, as long as the correct size $R_{\rm lobe}$ is taken for the spatially covered spectra, because the momentum is conserved along the outflow lobes. To test this assumption, a range of radii were chosen starting at the central position in each map and finishing at the border of the map. When taking a larger radius, more and more spectra are integrated, but $R_{\rm lobe}$ simultaneously increases. Nearby outflows from other sources are excluded beforehand. Panel 2 in Fig. \ref{allchoices} shows that the outflow force indeed remains constant as a function of radius. The outflow force of IRAS~4A increases significantly up to a factor 3 due to the inclusion of a ``bullet'' and high-velocity emission starting around 90$\arcsec$. 
Elias~29  decreases slightly at larger radii, due to the two broken receivers as shown in Fig. \ref{outflowmaps}, where parts of the lobes are missing. This figure proves that the outflow force can be analyzed with partial coverage as long as it is centered on the source position, but one has to be careful with emission from other sources, ``bullet'' emission and when regions of data are missing.

\subsubsection{Spectral binning}
Astronomical spectra are often rebinned to lower spectral resolution in order to decrease the rms, since lines of outflows have broad spectral profiles. When the outflow velocity limits are based on the rms level, binning to larger bin sizes may influence the outflow force significantly. This is investigated in panel 3 in Fig. \ref{allchoices}. 
The outflow force remains constant within a factor of 2 for Elias~29, IRS~54 and IRAS~4A. Inspection of the mass as function of bin size (not shown here) shows a decrease of the integrated mass, but this is apparently compensated by the higher outer velocities. On the other hand, HH~46 shows a significant increase of outflow force. This is due to the low $S/N$ of this data set: the lower rms moves $\varv_{\rm max}$ rapidly outwards while the inner velocity remains almost constant. Therefore, one has to be careful with binning spectra when using an rms cut-off for the velocities, since the choice of binning may further increase the outflow force significantly. The intensity-weighted methods, such as M7, are less affected by rebinning. 

\subsubsection{Cut-off level}
The rms level determines the outer velocity and therefore the outflow force depends on the quality of the data. In order to investigate this, the observed high quality spectra are convolved with an artificial noise signal and the velocity limits are redetermined with the new rms. The resulting outflow force as function of new rms is shown in panel 4 of Fig. \ref{allchoices}: the outflow force can be underestimated by a factor of 5--10 for large noise levels ($S/N<5$ for the wing emission). The outflow force of HH~46, which already has a large noise level to start with, is likely to be underestimated.

In some studies, the rms cut-off is not taken at 1$\sigma$, but rather 3$\sigma$ or even 10$\sigma$ to determine the outer velocities.  The effect of taking a cut-off at multiple $\sigma$ is investigated in panel 5. The outflow force decreases slowly when taking a stricter cut-off but the overall effect remains within a factor of 3-4, except for the spectra of HH~46 with low $S/N$.

\subsubsection{$\varv_{\rm out}$: ratio cut-off}
Due to the disadvantages of using an rms cut-off described above, another method is sometimes applied. The cut-off limit is defined as e.g. 1\% of the intensity at the inner velocity limit of the outflow wing (DC07). In that case, the cut-off level depends on the brightness of the source itself, rather than the quality of the observations. The difference between these two methods is investigated in Panel 6 of Fig. \ref{allchoices}, where the outflow force is normalized to the value found by the rms cut-off. The outflow force is likely to be underestimated for a ratio smaller than 50. For the larger ratios, the outflow force remains constant, which means that the outflow wing is sufficiently steep: the outer velocity is not affected by this small change in the cut-off level. 

\subsubsection{Summary}
Observational properties and choices in the analysis procedure can affect the outflow force up to a factor of 4--5. Small changes in the inner velocity limit, the rms level, the cut-off significance and binning do not change the outflow force more than a factor of 2. It is demonstrated that the spatial coverage in principle does not influence the outflow force, indicating that this is a conserved quantity. 
Also, choosing a percentage of the emission at the inner velocity limit for the cut-off does not in principle change the outflow force, as long as the percentage is small enough. The most important factor to consider is the $S/N$, as shown in the results of HH~46 with a higher noise level. A strong dependence of the outflow force on the noise level (Fig. \ref{allchoices}) implies that the methods described cannot always be applied directly. 
When combining the results from different studies, using different methods, assumptions and/or data quality, scatter up to a factor of 5 can be expected.

\onecolumn

\section{Outflow parameters from other studies}
\begin{table}[!h]
\caption{Outflow and evolutionary parameters from other studies used in evolutionary trends.}
\label{trendparameters}
\centering
\begin{tabular}{lcccccc}
\hline
\hline
Name&Class&$F_{\rm CO}$&Ref. ($F_{\rm CO}$)&$L_{\rm bol}$&$M_{\rm env}$&Ref. ($L_{\rm bol}$,$M_{\rm env}$)\\
&&($M_{\odot}$ km s$^{-1}$&&($L_{\odot}$)&(10$^{-2} M_{\odot}$)&\\
&&yr$^{-1}$)&&&&\\
\hline
L1448-C&0&1.7($-$4)&1&9.0&390&2\\
L1455M-FIR&0&3.5($-$4)&1&16&-&1\\
HH7-11&I&1.2($-$3)&1&41.8&950&2\\
IRAS3282&0&5.6($-$3)&1&3.0&-&1\\
RNO43-FIR&0&1.2($-$3)&1&12&-&1\\
VLA1623&0&5.0($-$5)&1&0.27&111&3\\
IRAS16293&0&3.2($-$4)&1&16&3.8&4,5\\
L723&0&5.0($-$5)&1&3.6&130&2\\
B335&0&3.6($-$5)&1&3.3&120&2\\

L1489 IRS&I&1.4($-$5)&6&3.7&2.0&6\\
T Tau&I&7.3($-$5)&6&25.5&2.9&6\\
Haro 6-10&I&9.4($-$6)&6&6.98&0.40&6\\
L1551 IRS5&I&2.2($-$3)&6&21.9&26&6\\
L1535 IRS&I&1.9($-$7)&6&0.70&10&6\\
TMR 1&I&3.7($-$6)&6&3.8&20&2\\
TMC 1A&I&8.7($-$6)&6&2.7&20&2\\
L1527 IRS&0&1.6($-$4)&6&1.9&90&2\\
TMC 1&I&2.6($-$6)&6&0.9&20&2\\

Ced110IRS4&I&4.6($-$5)&7&0.8&20&2\\
CrAIRAS32&I&1.8($-$4)&7&3.4&80&7\\
HH100&I&$<$5.3($-$3)&7&17.7&810&2\\
RCrAIRS7&I&$<$4.9($-$2)&7&20&630&7\\
\hline
\end{tabular}
\tablebib{
(1) CB92, (2) \citet{kristensen2012}, (3) \citet{kempen2009}, (4) \citet{evans2009}, (5) \citet{jorgensen2008}, (6) \citet{hogerheijde1998}, (7) \citet{kempen2009outflow}
}
\end{table}

\section{Spectra overview}
\begin{figure*}[!ht]
\subfloat{\includegraphics[scale=0.5]{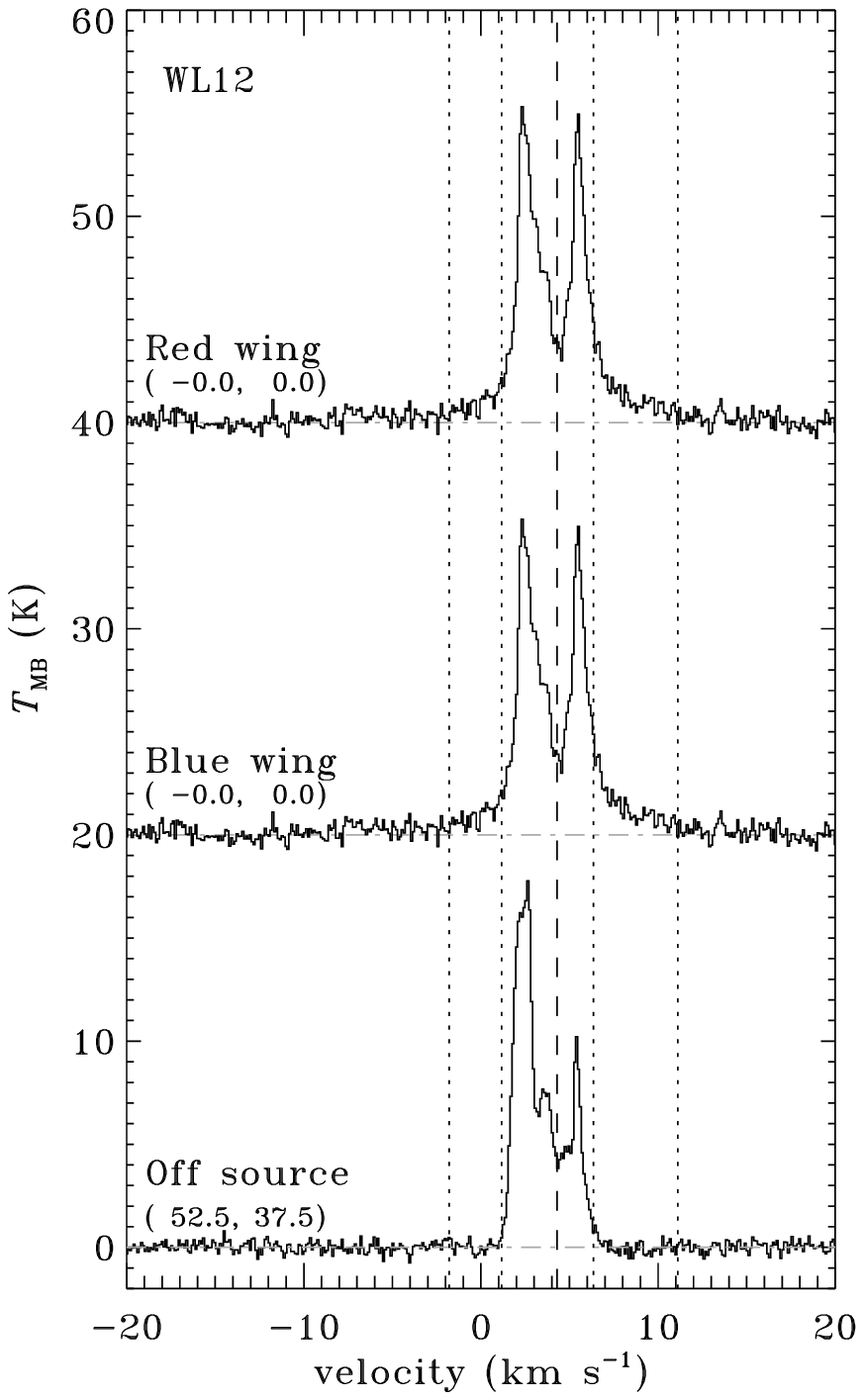}}
\subfloat{\includegraphics[scale=0.5]{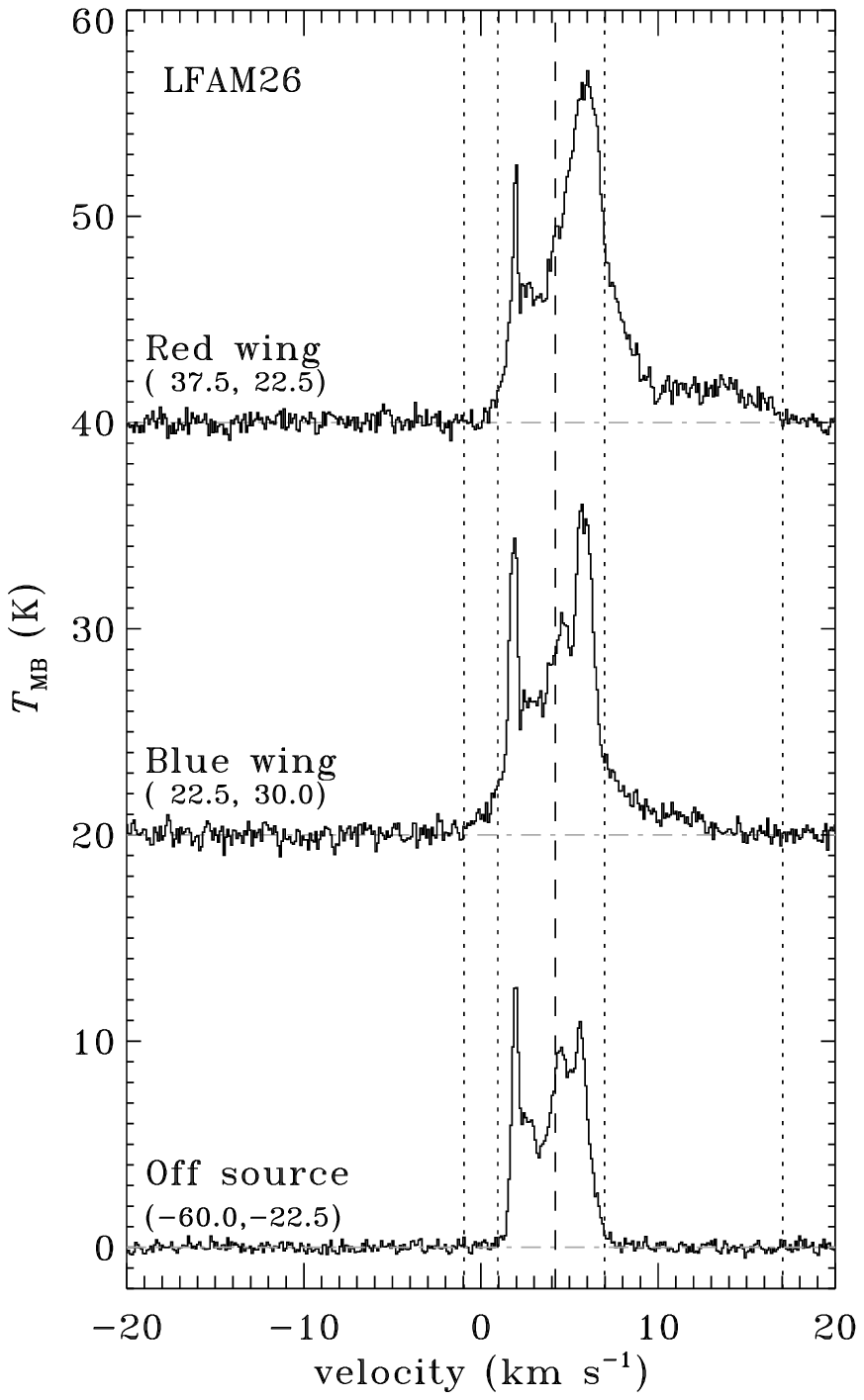}}
\subfloat{\includegraphics[scale=0.5]{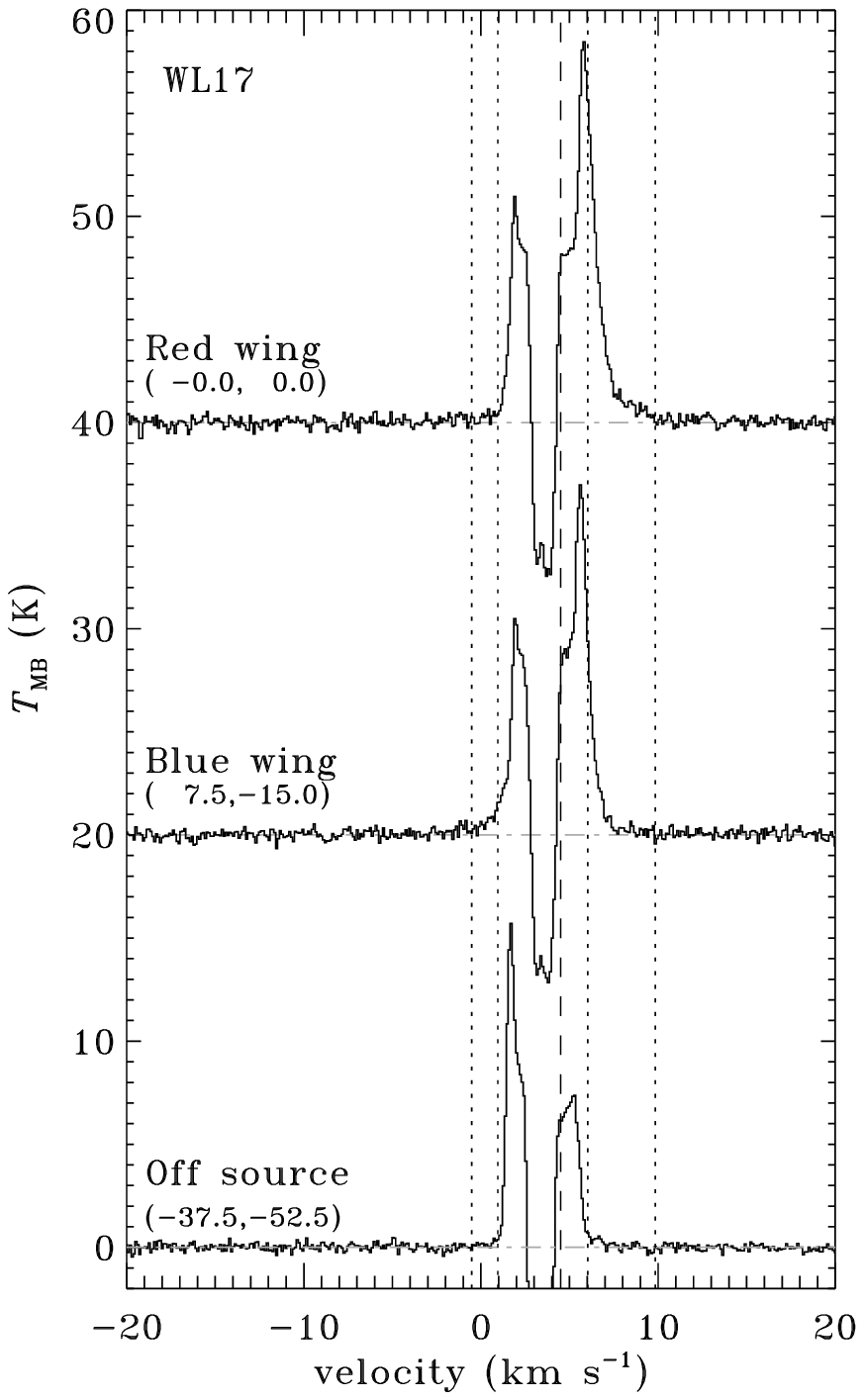}}\\
\captcont{Overview of relevant outflow spectra. Each diagram shows from top to bottom the strongest red wing spectrum, the strongest blue wing spectrum and an off-source spectrum. $\varv_{\rm source}$ (dashed line) and the integration limits (dotted lines) are indicated. A dash-dotted line indicates the baseline. Positions of each spectrum are indicated in parentheses.}
\end{figure*}

\begin{figure*}[!ht]
\subfloat{\includegraphics[scale=0.5]{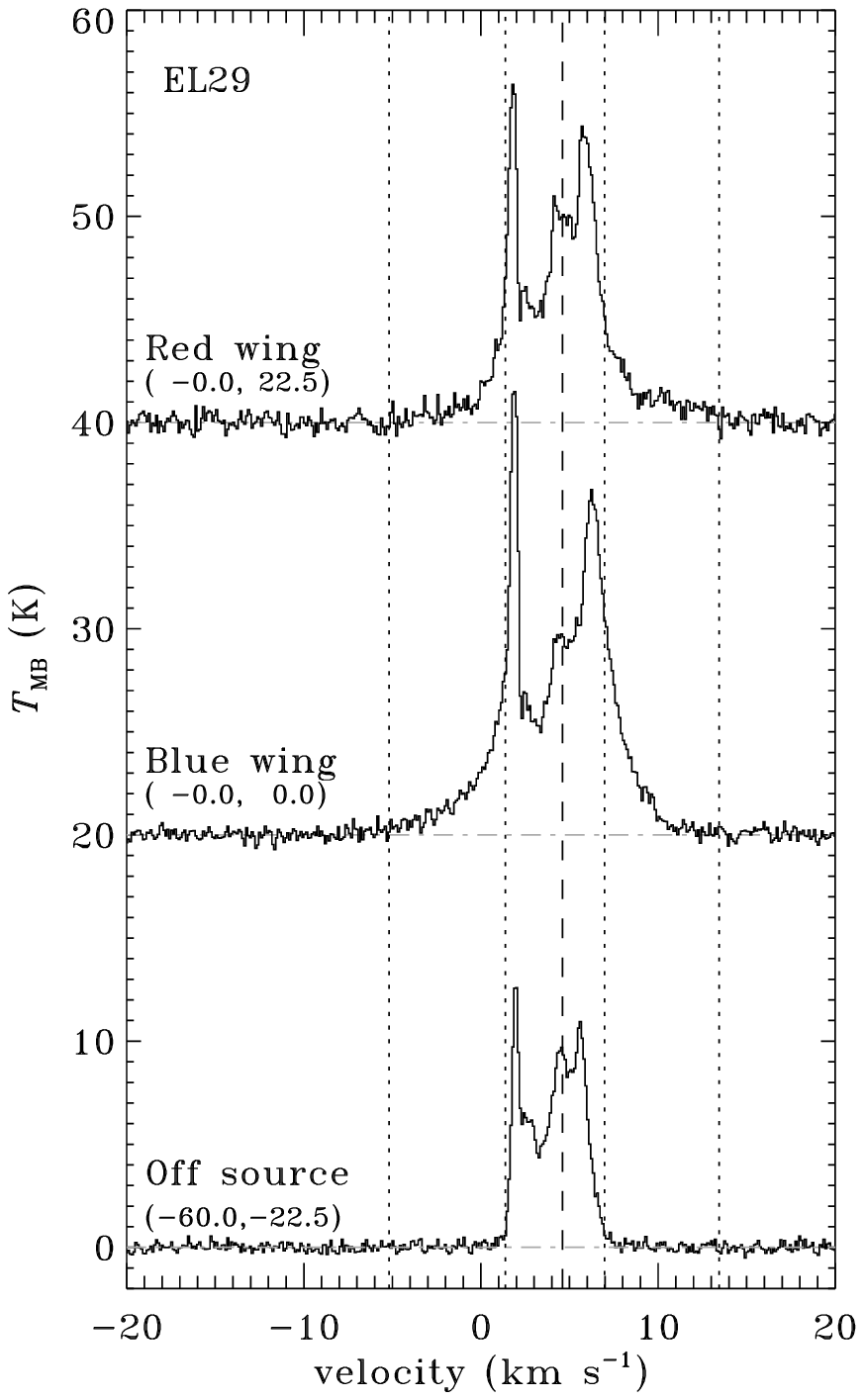}}
\subfloat{\includegraphics[scale=0.5]{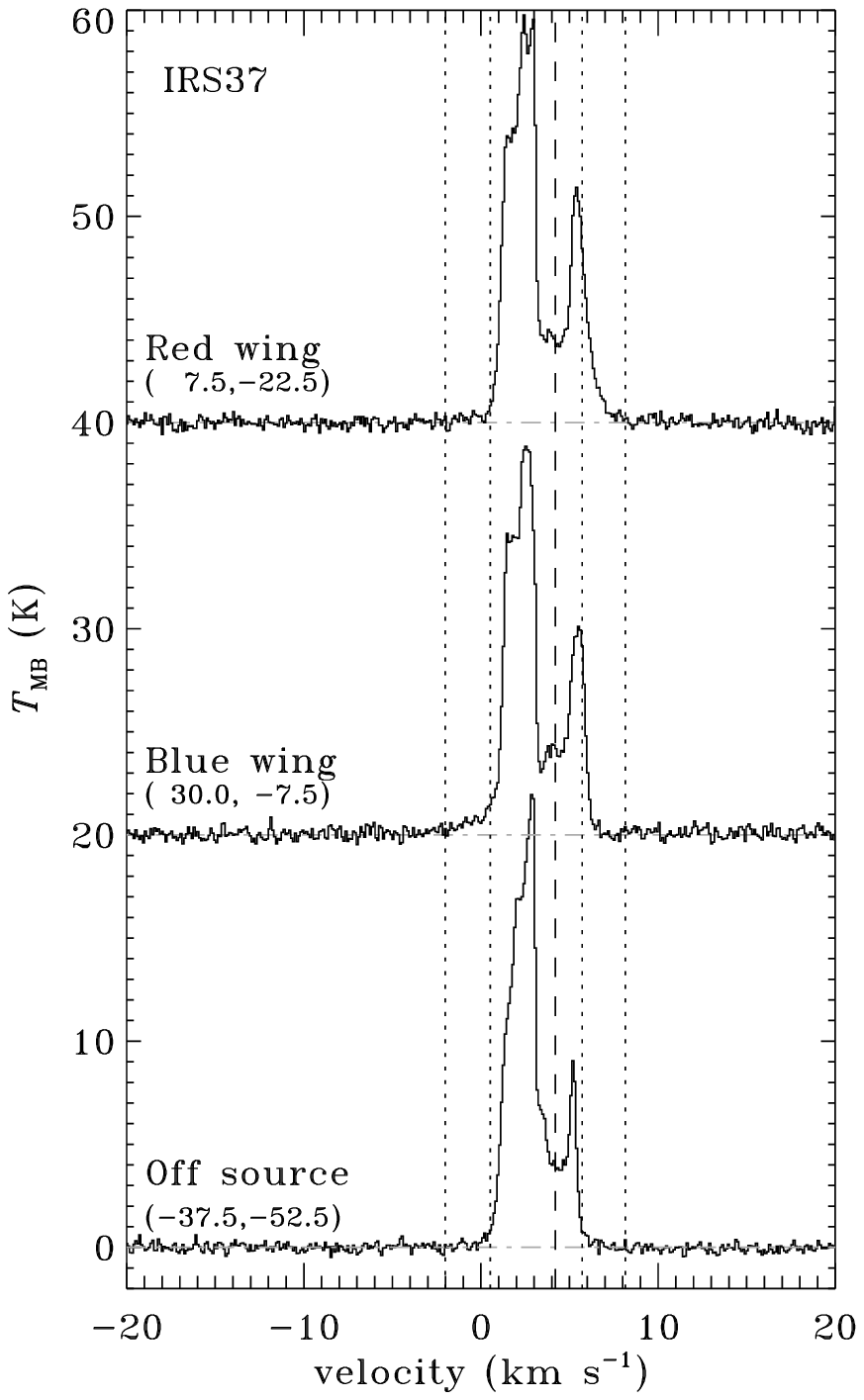}}
\subfloat{\includegraphics[scale=0.5]{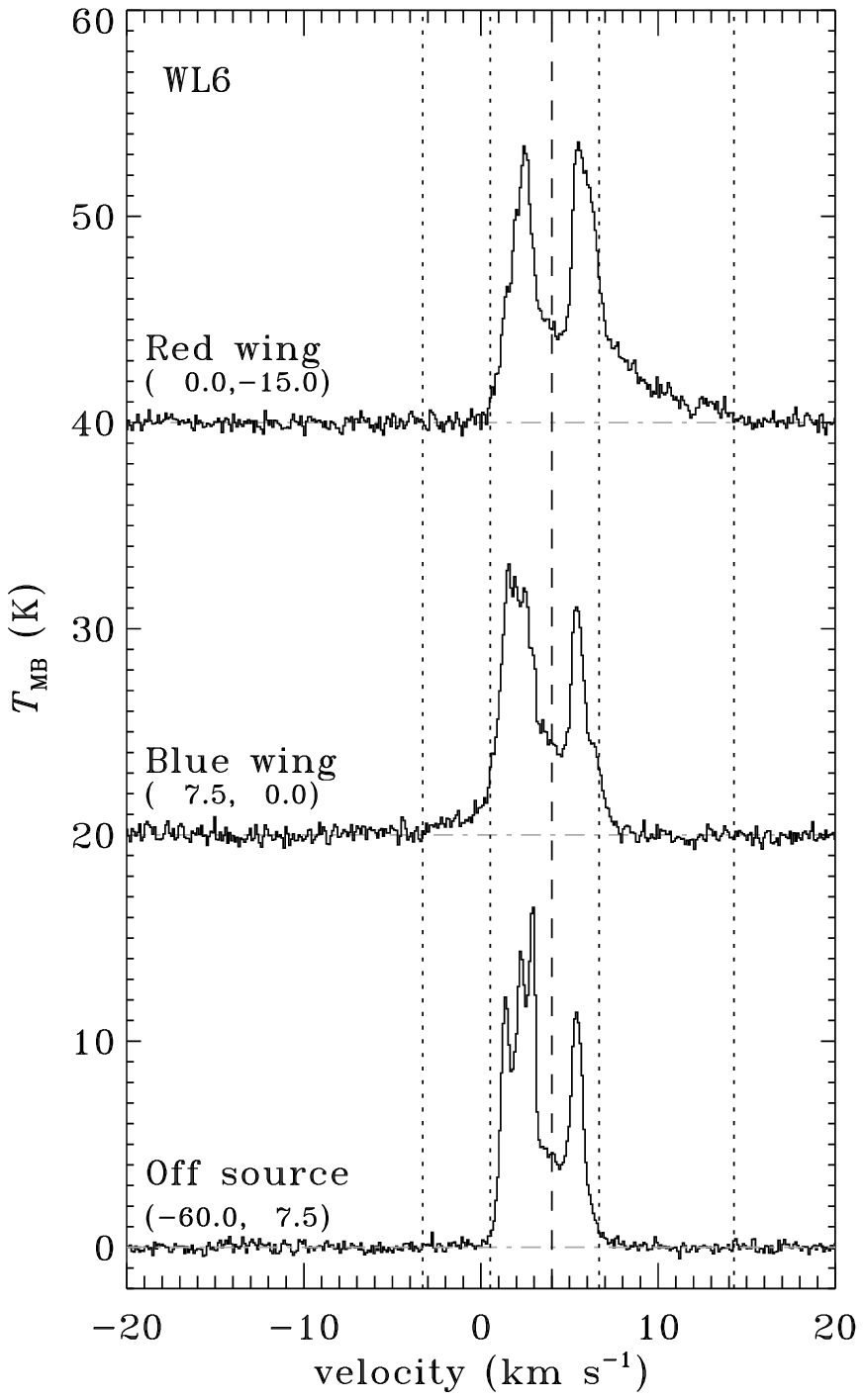}}\\
\subfloat{\includegraphics[scale=0.5]{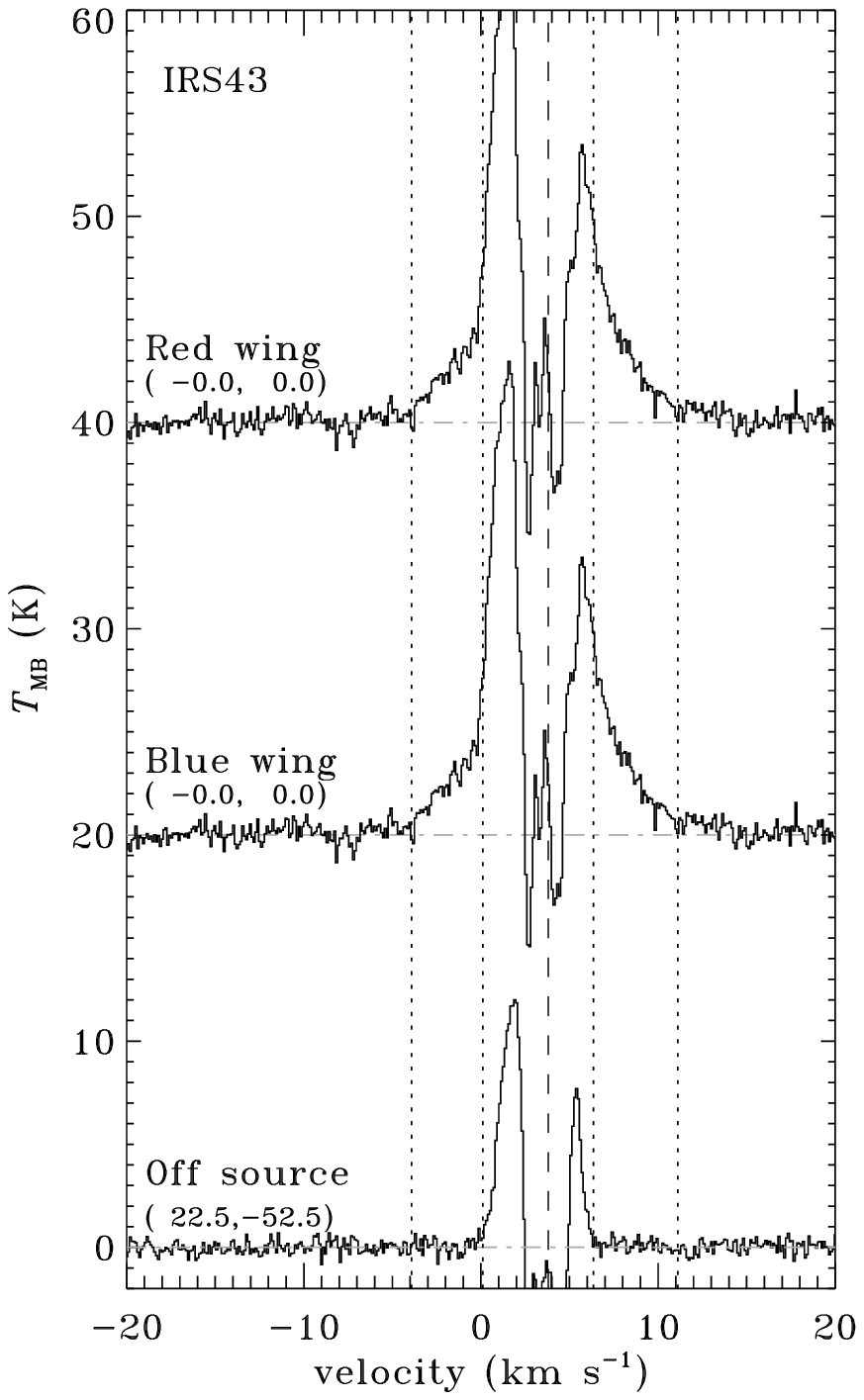}}
\subfloat{\includegraphics[scale=0.5]{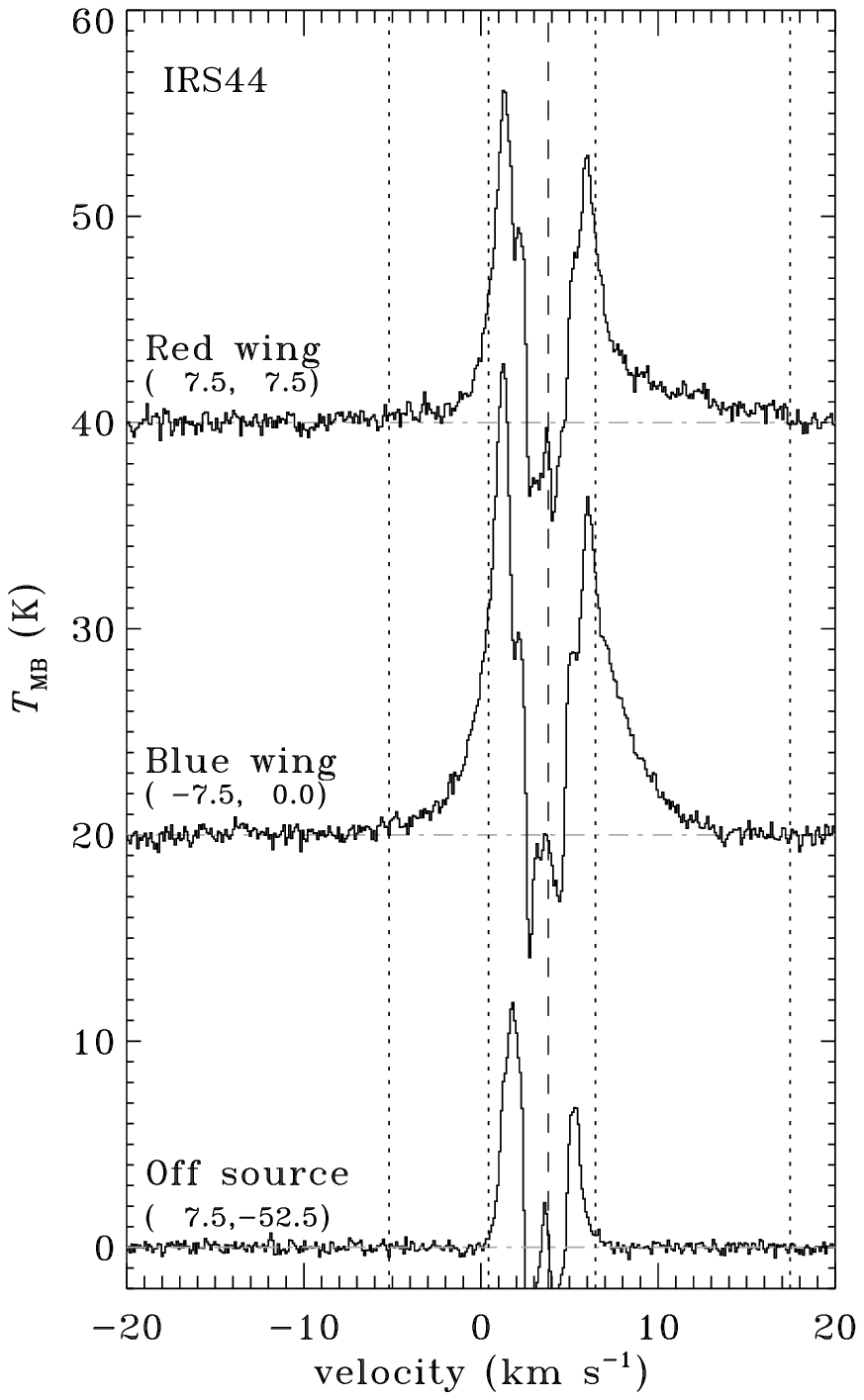}}
\subfloat{\includegraphics[scale=0.5]{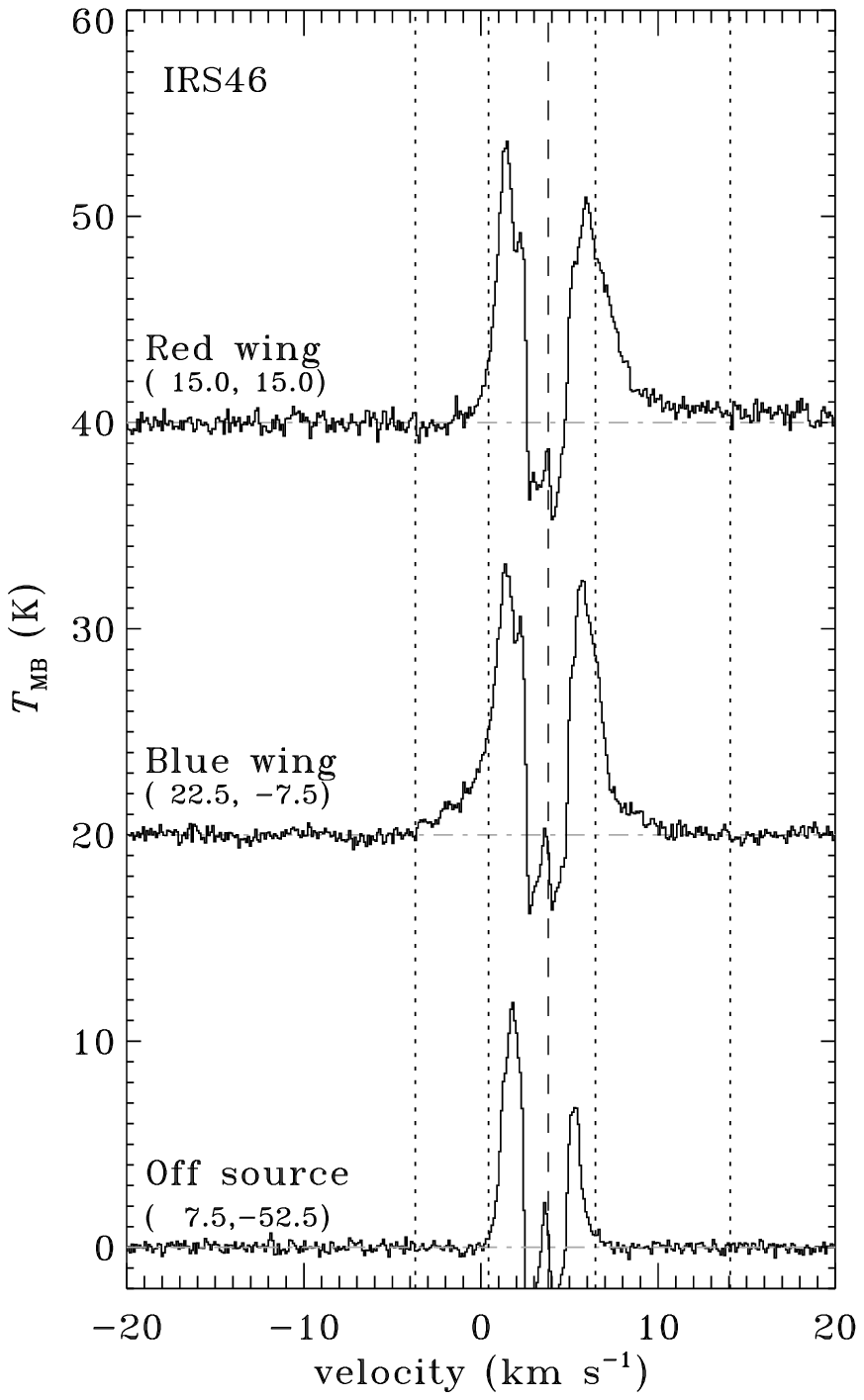}}\\
\subfloat{\includegraphics[scale=0.5]{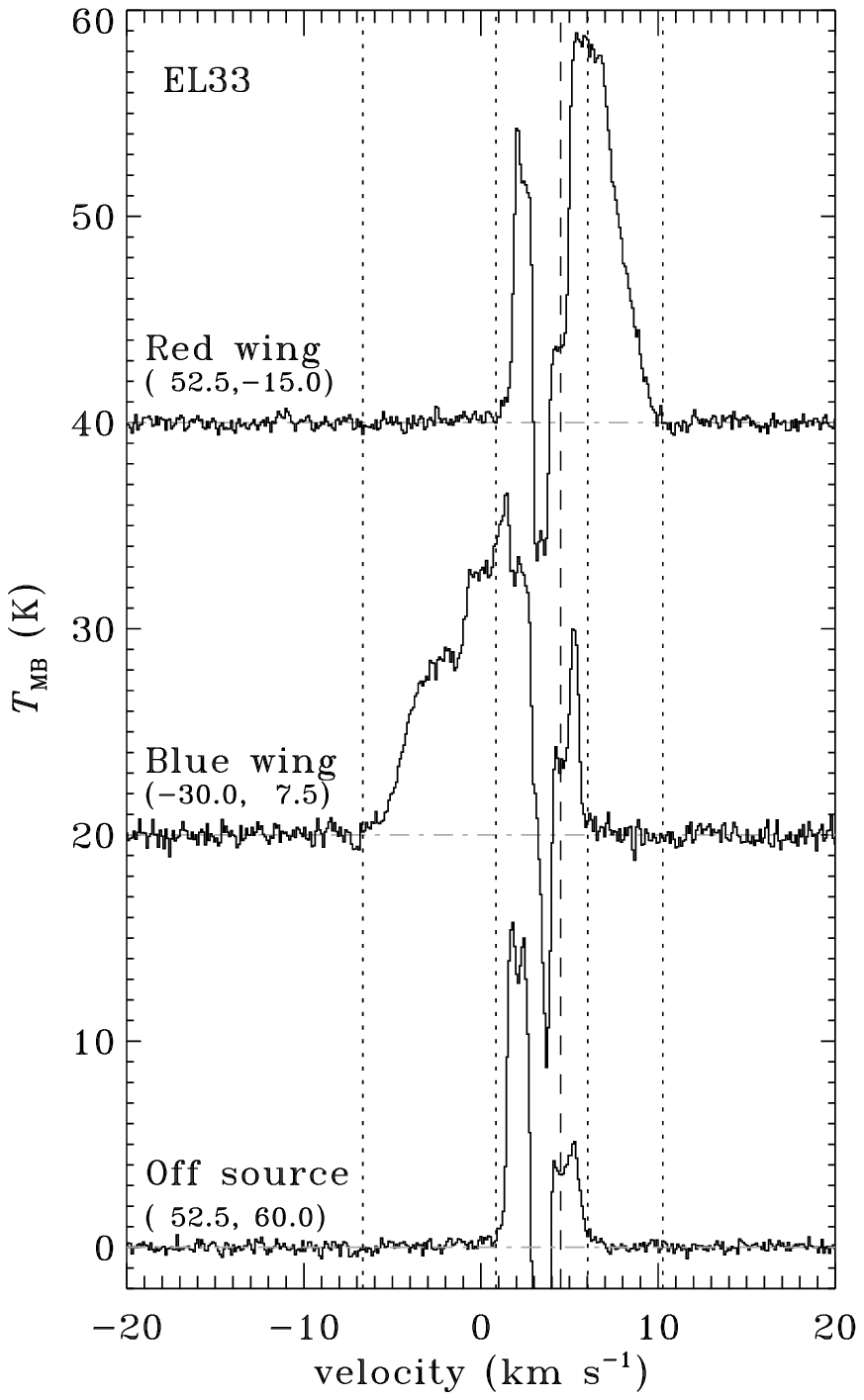}}
\subfloat{\includegraphics[scale=0.5]{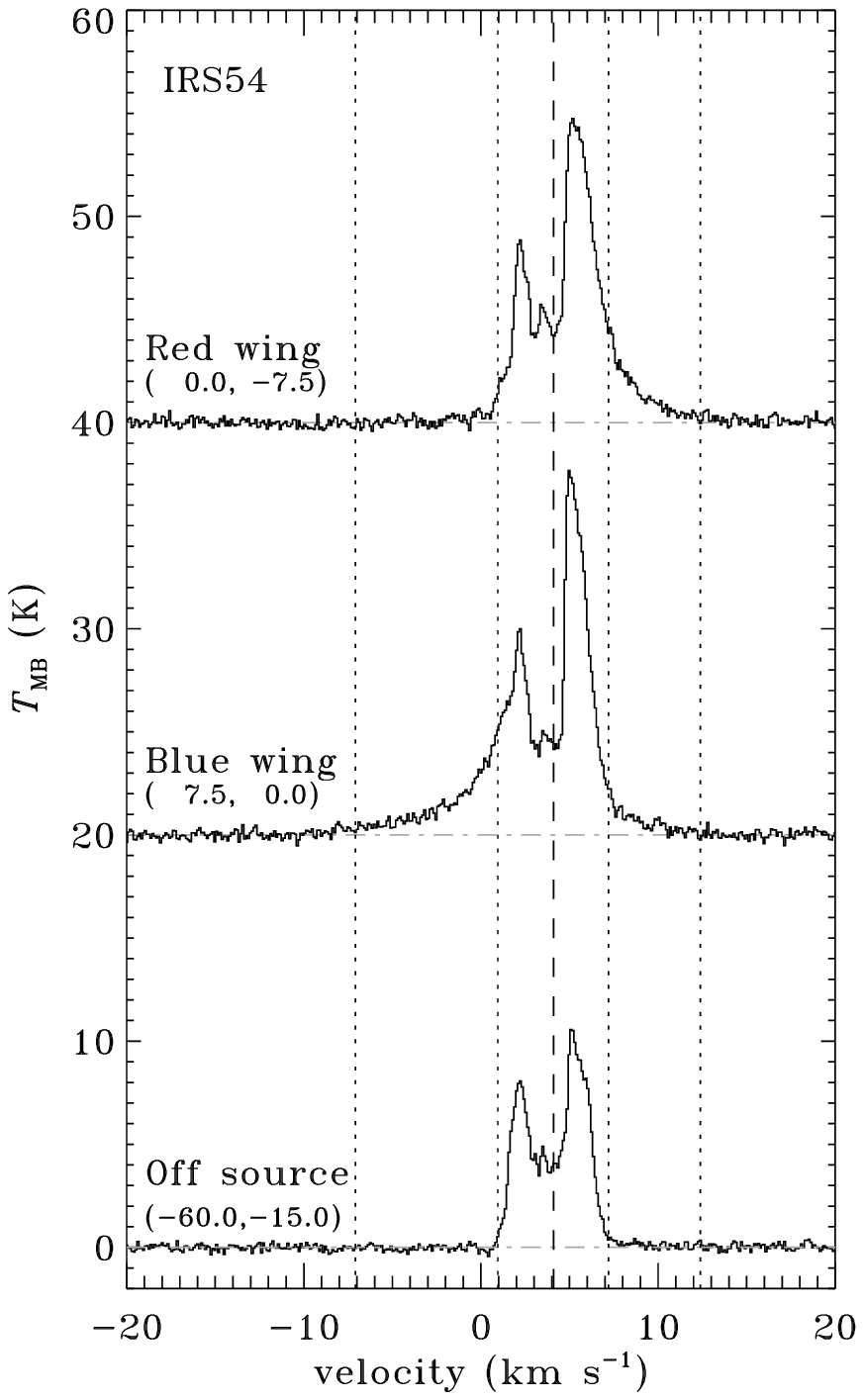}}
\subfloat{\includegraphics[scale=0.5]{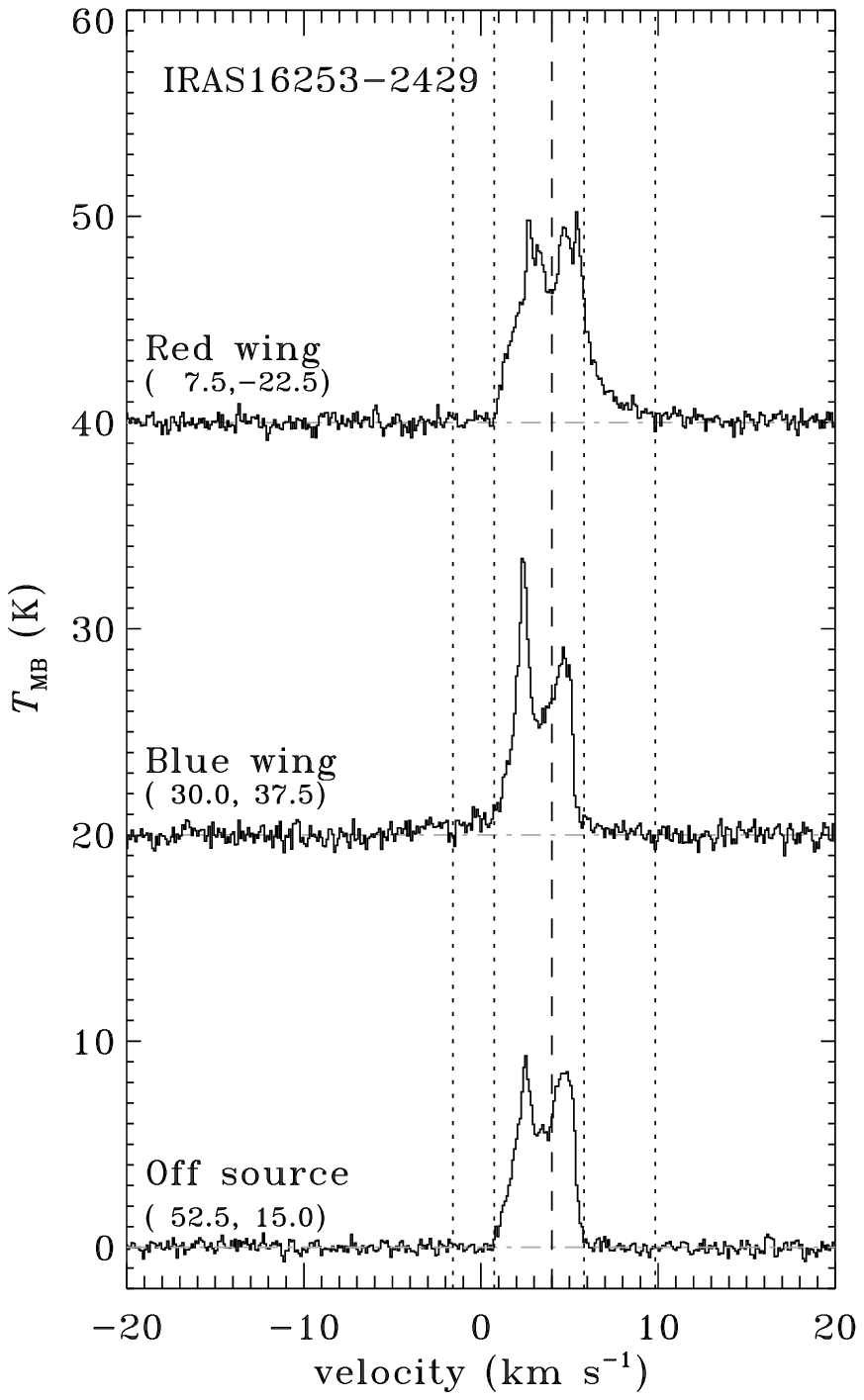}}\\
\captcont{Overview of relevant outflow spectra - Continued}
\end{figure*}

\begin{figure*}[!ht]
\subfloat{\includegraphics[scale=0.5]{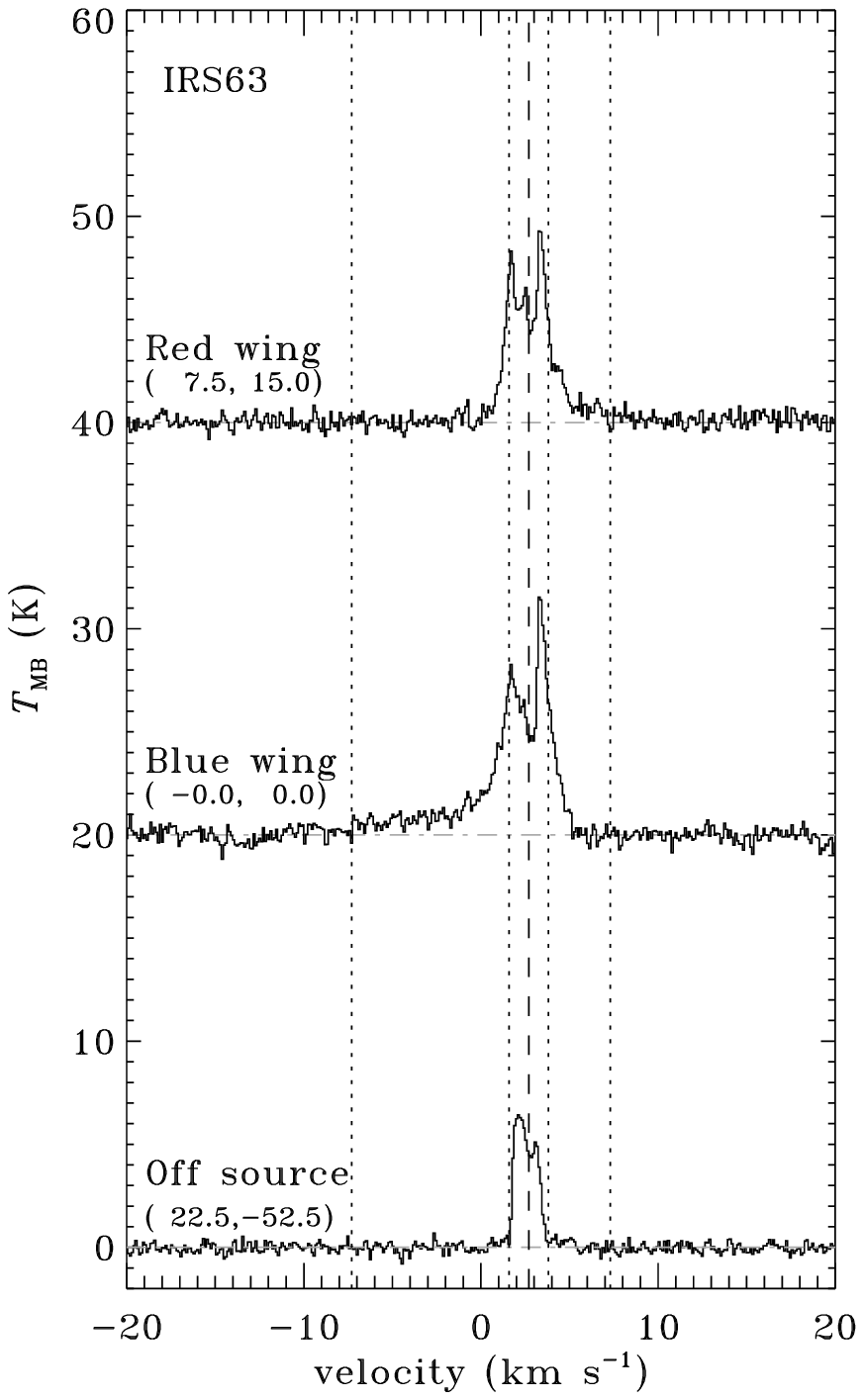}}
\subfloat{\includegraphics[scale=0.5]{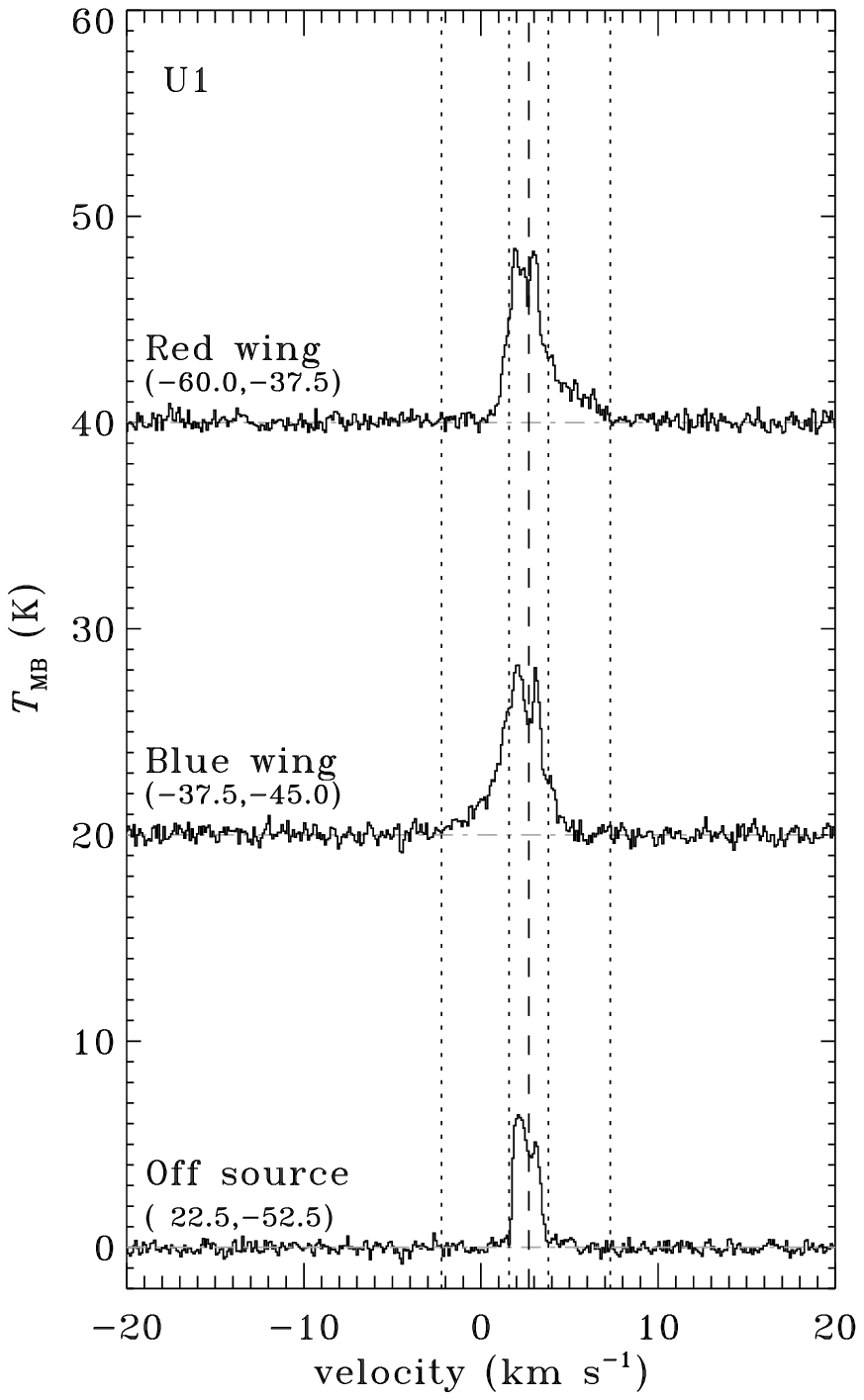}}
\subfloat{\includegraphics[scale=0.5]{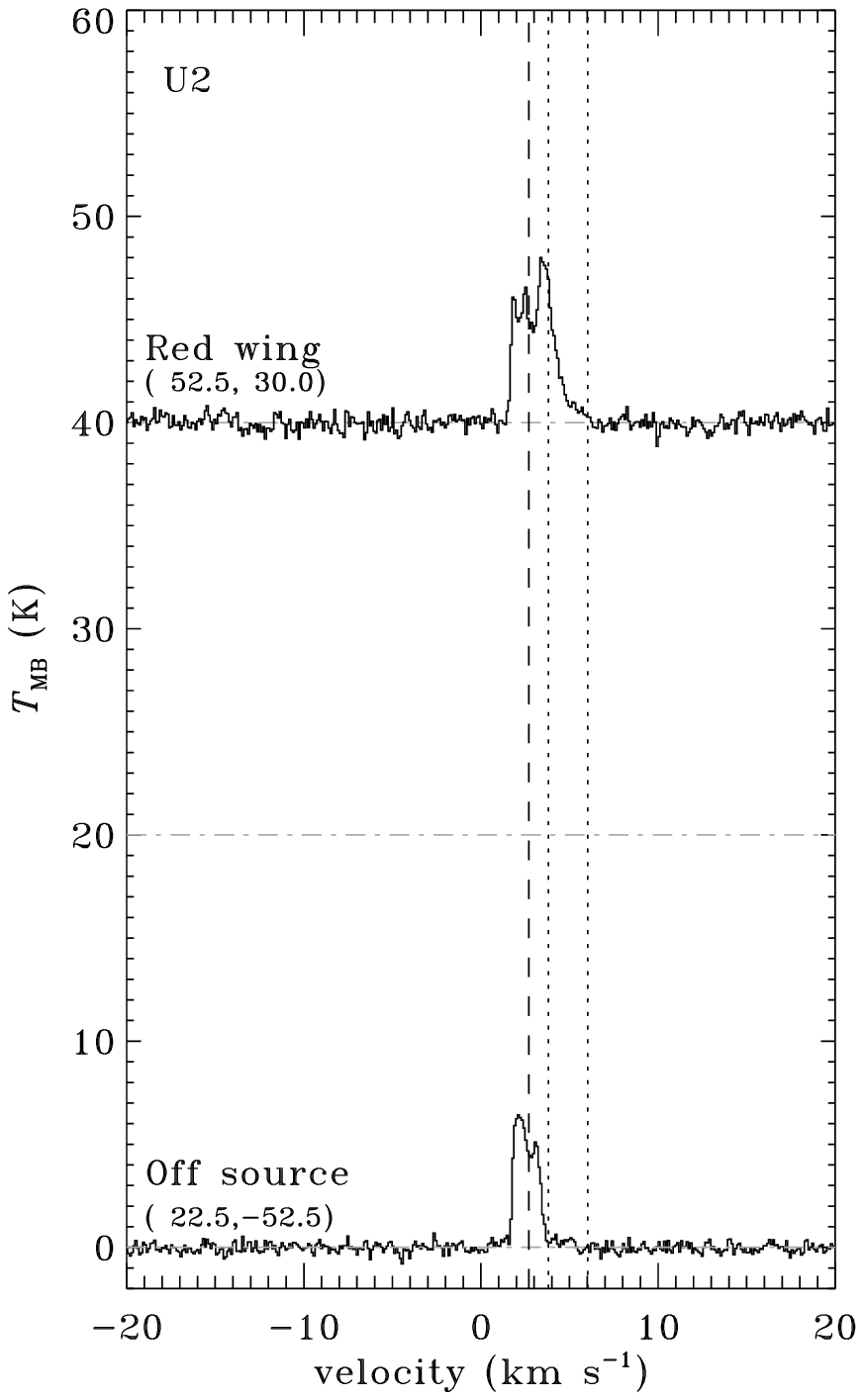}}\\
\subfloat{\includegraphics[scale=0.5]{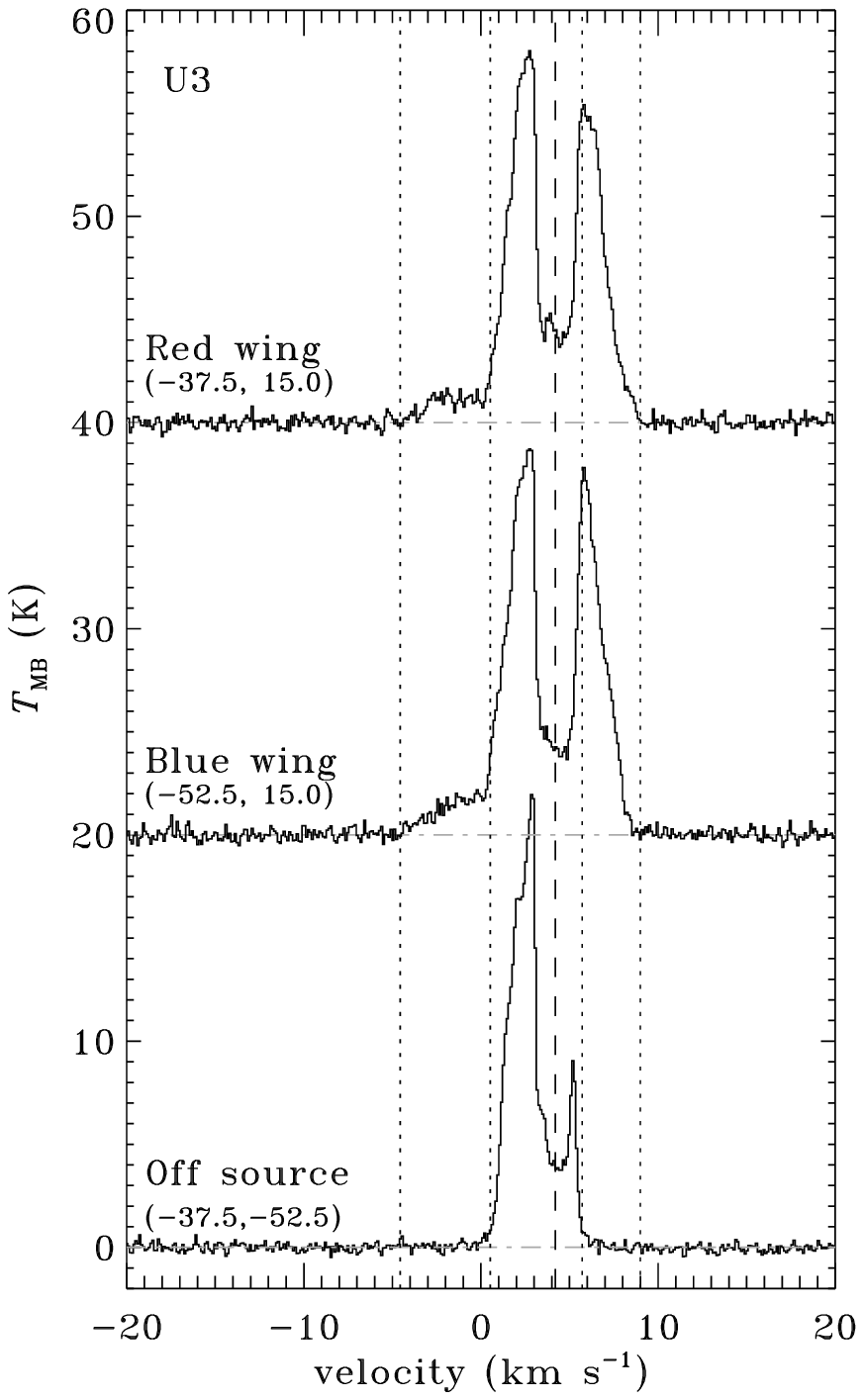}}
\subfloat{\includegraphics[scale=0.5]{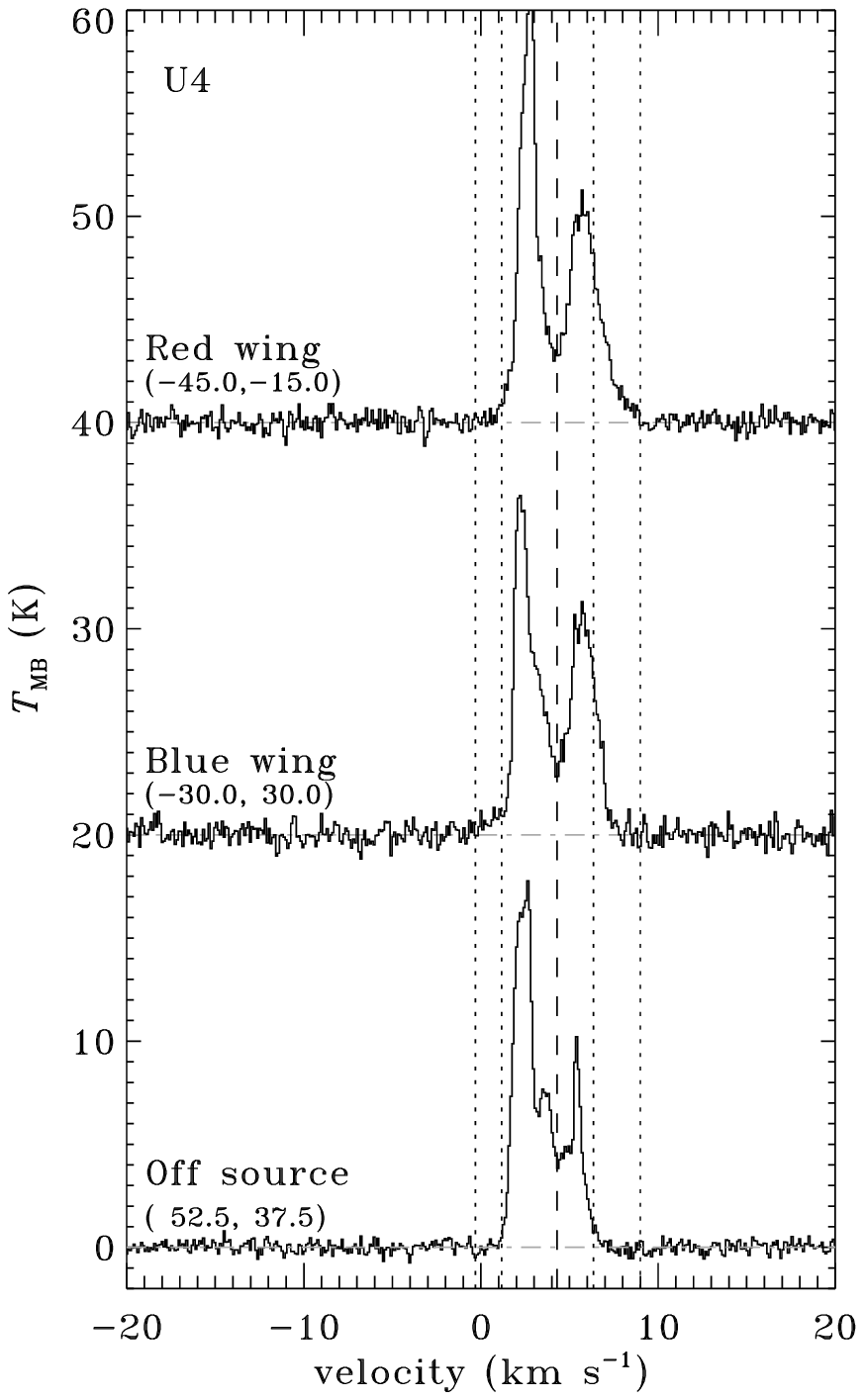}}
\subfloat{\includegraphics[scale=0.5]{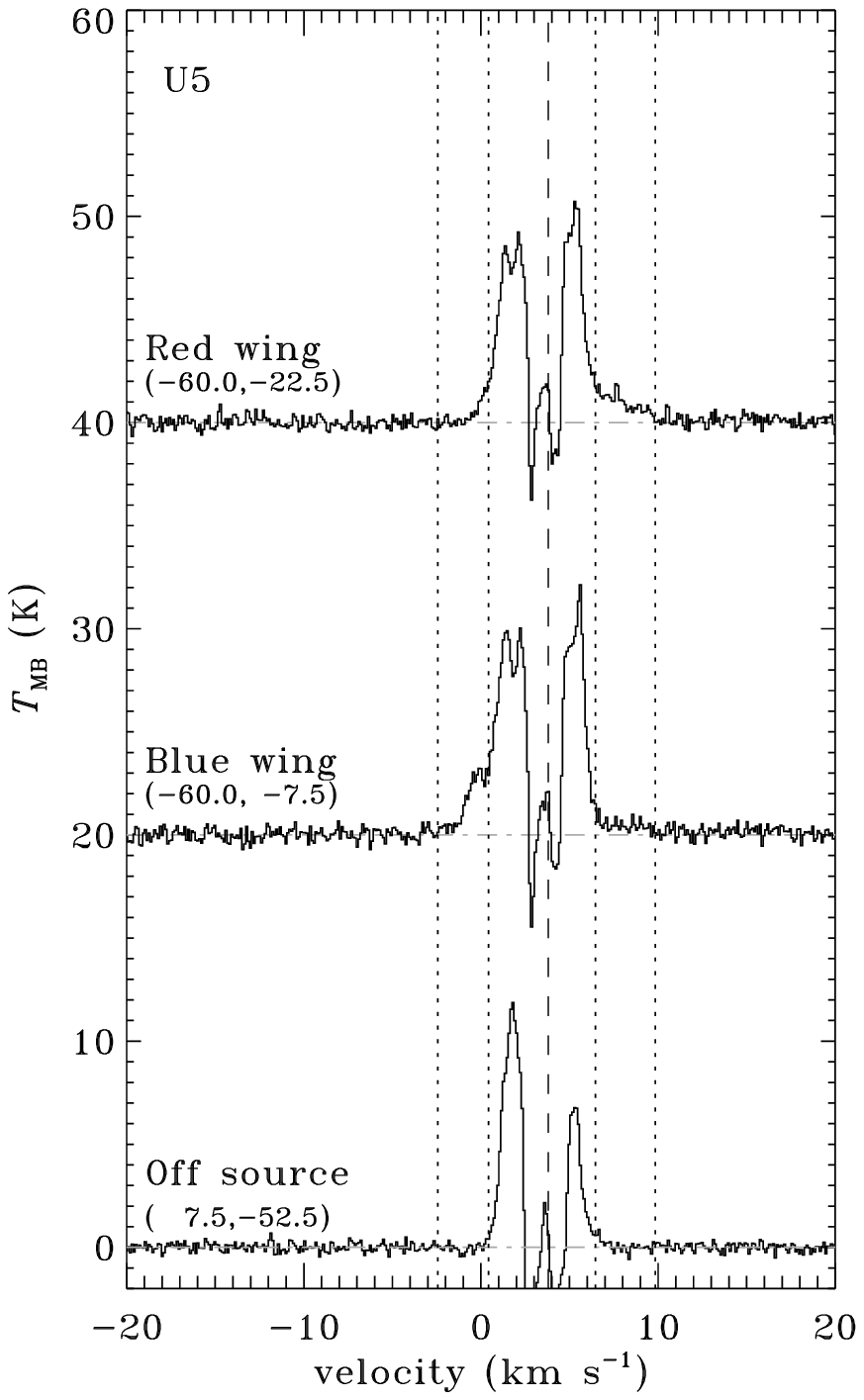}}\\
\subfloat{\includegraphics[scale=0.5]{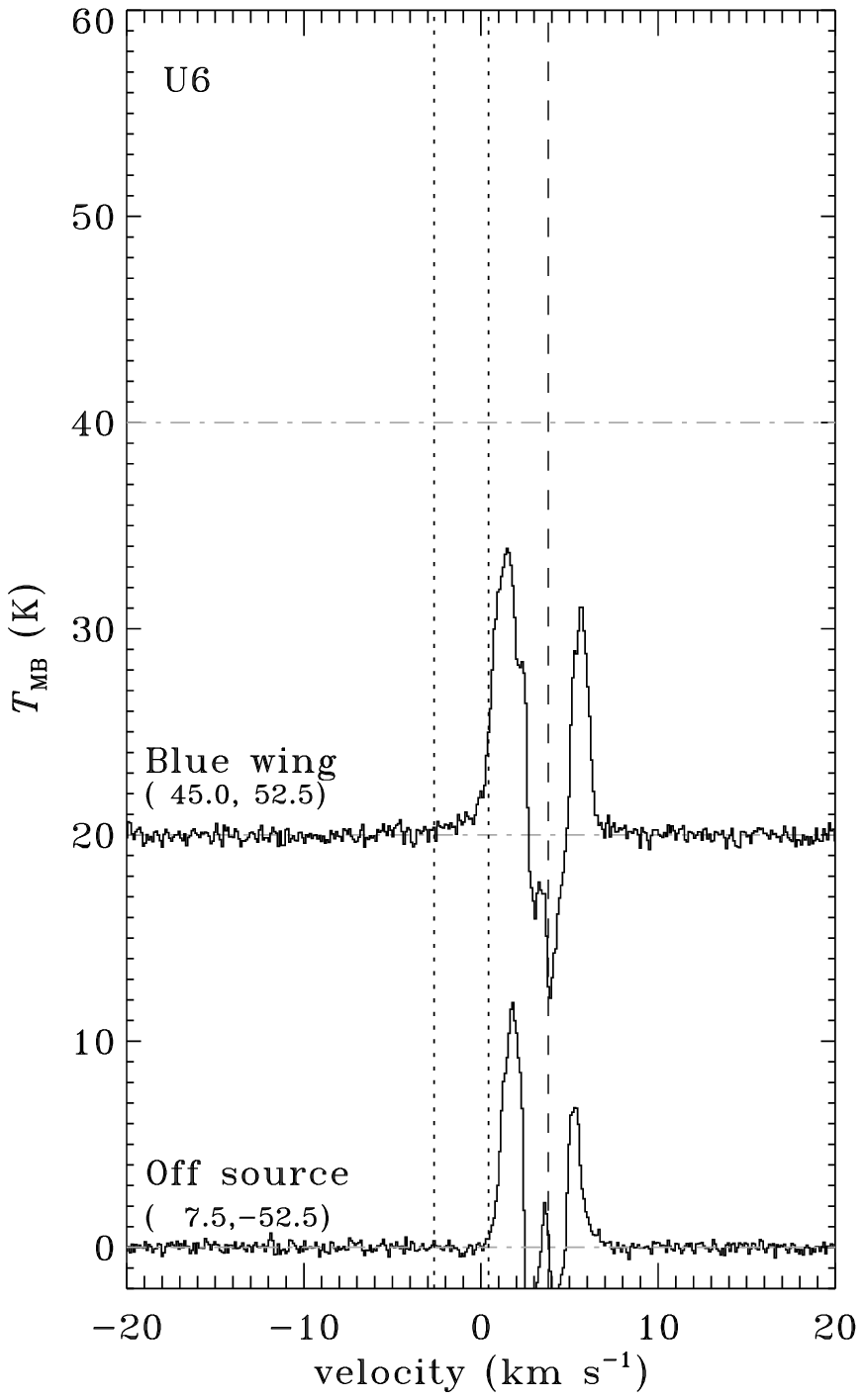}}
\caption{Overview of relevant outflow spectra - Continued}
\label{overviewspect}
\end{figure*}

\clearpage

\begin{table*}[ht]
\caption{Integration limits of the $^{12}$CO $J$=3--2 line wings.}
\label{limits}
\centering
\begin{tabular}{lccccc}
\hline
\hline
&&\multicolumn{2}{c}{Blue lobe}&\multicolumn{2}{c}{Red lobe}\\
\cline{3-4}\cline{5-6}
Name&$\varv_{\rm source}$&$\varv_{\rm out}$&$\varv_{\rm in}$&$\varv_{\rm out}$&$\varv_{\rm in}$\\
&(km s$^{-1}$)&(km s$^{-1}$)&(km s$^{-1}$)&(km s$^{-1}$)&(km s$^{-1}$)\\
\hline
WL~12&4.3&-1.8&1.2&6.3&11.1\\
LFAM~26&4.2&-0.7&1.0&6.7&11.5\\
LFAM~26&4.2&-1.0&1.0&7.0&17.0\\
WL~17&4.5&-0.5&1.0&6.0&9.8\\
EL~29&4.6&-5.2&1.4&7.0&13.4\\
IRS~37&4.2&-2.0&0.5&5.7&8.1\\
WL~6&4.0&-3.3&0.5&6.7&14.3\\
IRS~43&3.8&-3.9&0.1&6.3&11.1\\
IRS~44&3.8&-5.2&0.4&6.5&17.5\\
IRS~46&3.8&-3.7&0.4&6.5&14.1\\
EL~33&4.5&-6.7&0.8&6.0&10.3\\
IRS~54&4.1&-7.1&1.0&7.2&12.4\\
IRAS~16253-2429&4.0&-1.6&0.7&5.8&9.8\\
IRS~63&2.7&-7.3&1.6&3.8&7.3\\
U1&2.7&-2.2&1.6&3.8&7.3\\
U2&2.7&0.0&1.6&3.8&6.0\\
U3&4.2&-4.6&0.5&5.7&9.0\\
U4&4.3&-0.3&1.2&6.3&9.0\\
U5&3.8&-2.4&0.4&6.5&9.8\\
U6&3.8&-2.6&0.4&6.5&8.6\\
\hline
\end{tabular}
\end{table*}

\twocolumn
\clearpage

\begin{figure}[h]
\begin{center}
\includegraphics[scale=0.6]{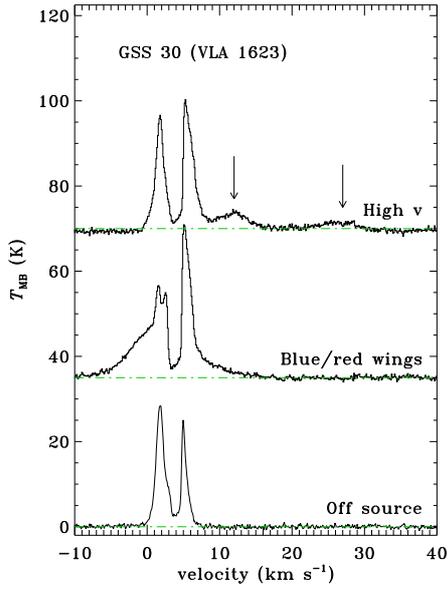}
\caption{Spectrum with extremely high velocity (EHV) emission in the GSS~30 map as indicated by arrows, originating from the Class 0 source VLA~1623. Any outflow emission from GSS~30 itself can not be disentangled from VLA~1623 in the present observations. From top to bottom: one high velocity position ($-$45$\arcsec$, 0$\arcsec$), one wing position ($-$30$\arcsec$, $-$22$\farcs$5) and one off-source position  (37$\farcs$5, 22$\farcs$5). The baselines are marked by dashed lines. The high-velocity bullet at 28 km s$^{-1}$ has not been identified in earlier studies of VLA~1623.}
\label{GSS30highvelspec}
\end{center}
\end{figure}

\begin{figure}[h]
\begin{center}
\includegraphics[scale=0.6]{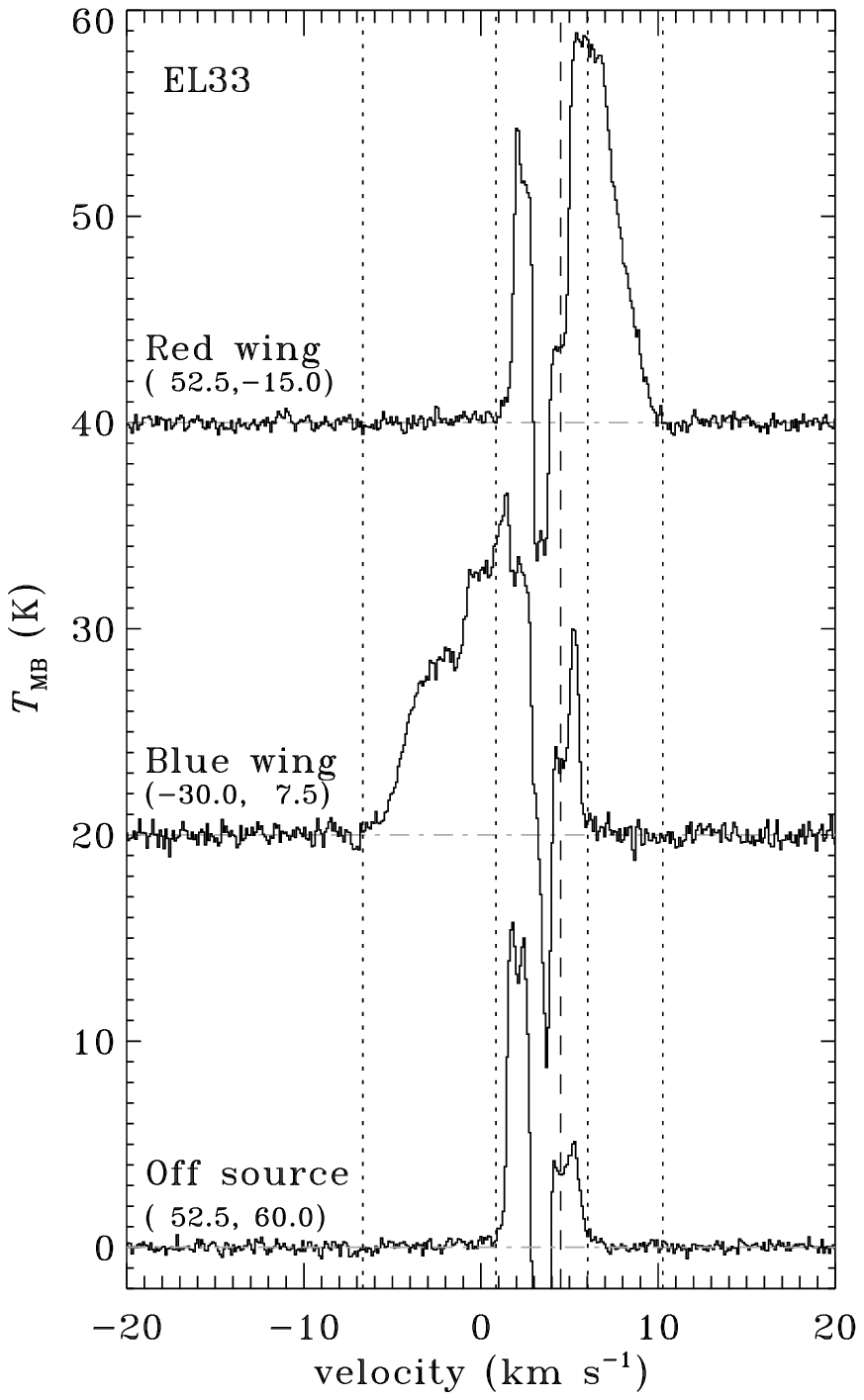}
\caption{Outflow spectra of Elias 32/33, showing a red wing spectrum (top), blue wing spectrum (middle) and off-source spectrum (bottom). $\varv_{\rm source}$ (dashed line) and the integration limits (dotted lines) are indicated. A dash-dotted line indicates the baseline. Positions of each spectrum are indicated in brackets. Especially the blue wing shows a double structure, suggesting blending of outflow emission of two separate sources.}
\label{elias33}
\end{center}
\end{figure}

\end{document}